\documentclass[preprints,review,accept,oneauthor,10pt,a4paper]{mdpi}

\firstpage{1}
\pubvolume{xx}
\issuenum{1}
\articlenumber{1}
\pubyear{2018}
\copyrightyear{2018}
\externaleditor{Academic Editor: name}
\history{Received: 29 August 2018; Accepted: 9 November 2018; Published: 15 November 2018}

\pdfoutput=1

\usepackage{amsmath}
\usepackage{amssymb}

\Title{Conformal Symmetry in Field Theory\\ and in Quantum Gravity}

\Author{Les\l{}aw Rachwa\l{} $^{1}$\orcidA{}}

\AuthorNames{Leslaw Rachwal}

\address{%
$^{1}$ \quad Instituto de F\'{i}sica, Universidade de Bras\'{i}lia,
70910-900, Bras\'{i}lia, DF, Brazil; grzerach@gmail.com}

\abstract{Conformal symmetry always played an important role in field theory (both quantum
and classical) and in gravity. We present construction of quantum conformal gravity and discuss
its features regarding scattering amplitudes and quantum effective action. First, the long and
complicated story of UV-divergences is recalled. With the development of UV-finite higher derivative
(or non-local) gravitational theory, all problems with infinities and spacetime singularities are solved. 
Moreover, the non-local quantum conformal theory reveals itself to be ghost-free,
so the unitarity of the theory is safe. After the construction of UV-finite theory, we focused on
making it manifestly conformally invariant using the dilaton trick. We also argue that in this class of
theories conformal anomaly vanishes by fine-tuning the couplings. As applications
of this theory, the constraints of the conformal symmetry on the form of the effective action and
on the scattering amplitudes are shown. We also remark about the preservation of the unitarity
bound for scattering. Finally, the old model of conformal supergravity by Fradkin and Tseytlin is
briefly presented.}


\begin{document}

\section{Introduction }

From the beginning of research on theories enjoying invariance under
local spacetime-dependent transformations, conformal symmetry played
a pivotal role---{first introduced} by Weyl related changes of meters
to measure distances (and also due to relativity changes of periods
of clocks to measure time intervals). Weyl thought of transformations
changing the scale and quite boldly he considered them not only in
global version (where the parameters of the transformations are constant),
but also in local (where the parameters depend both on space location
and time). Then, he also understood that just rescaling transformations
\begin{equation}
ds\to ds'=e^{\lambda}ds\label{eq: rescalings}
\end{equation}
 (which form abelian group) after making them local gives rise to full
conformal group of transformations, where only angles remain invariant,
but the sizes, magnitudes and scalar products between vectors change.
These primordial considerations were done within a set of ideas that
conformal transformations are the symmetries of fully relativistic
consistent classical theory of gravitation. However, Einstein beat
the Weyl theory with his famous second argument about radiating atoms
in various gravitational fields. With the triumph of Einsteinian relativistic
theory of gravitation (confirmed by the measurements done during the
solar eclipse in 1917), which includes only symmetries with respect
to diffeomorphism transformations, Weyl's conformal theory of gravitation
was relegated and treated as non-physical. However, the ideas of scale
transformations seeded by Weyl were revealed to be very fruitful in
the development of local gauge theories. The change was that the transformations
were done in the internal space, not in spacetime, and that they were
imaginary rescalings or simply complex phase transformations compared
to the ones in Eqn. (\ref{eq: rescalings}). This 
with London equations
started a revolution leading to $U(1)$ gauge theory known as electrodynamics
and even to non-abelian generalizations as embodied by Yang--Mills
(YM) theories. However, the origin of the German word ``gauge'' used
now so often in field theory is clear and undoubtedly points to different
real scales used for measurement of distances, like for example between
two rails in rail transport.

It seemed that in the classical world conformal symmetry had to play a
rather small role and was not used in relation to classical gravity.
The time for conformal methods in general relativity (GR) had to come later,
for example when the research on black holes and their spacetime
causal structure culminated in the golden age of classical gravitation
in the 1960s and 1970s. In a different vein, conformal invariance and
conformal symmetry started to become appreciated a lot also in the
era of modern quantum field theories. With the development of the
renormalization methods in perturbative quantum field theory models,
conformal symmetry became an indispensable tool for every particle
physicist. Moreover, the relation between Renormalization Group (RG)
flow and Conformal Field Theories (CFT) and conformal anomalies was
found to be even more tight due to the understanding of physics at
and near a fixed point of RG. The application of conformal considerations
in field theory was also beautifully unified with gravitational (spacetime)
considerations through AdS/CFT duality. Now, every modern
theoretician knows that conformal tools are both useful and enlightening
for investigations both in gravity and in field theory. In the advent
of bigger understanding of conformal symmetry, we could come back and
refresh interest in the original Weyl gravity, where the original conformal
symmetry is taken as a basis for the theory of gravitational interactions.
However, on the quantum level, we meet several serious obstacles. Actually,
one can find that there is a very special relation between quantum
physics and conformal symmetry, which is not enjoyed by any other
known symmetries of Nature, nor by other gauge symmetries used in
Nature.

To get effective field theories (EFT), we typically need to perform a very
abstract mathematical procedure called quantization. It takes a classical
action of a theory and enforces the rules of quantum mechanics. In
all perturbative approaches, the result of this artificial procedure
done only by theoreticians (because the Nature is already always quantum!)
is expressed as a perturbative series in some small parameter. Typically,
this is a Planck constant $\hbar$ multiplying some couplings of the
theory and this is equivalent to the loop expansion. It is necessary here to 
distinguish the two possible meanings of EFT: they appear
either as a low-energy effective description (in powers of a small length 
scale) or as a quantum effective description (in powers of Planck constant).
Now, the question
arises of how to find effective actions of a given model at any loop
order. By the knowledge and experience gained in some ordinary QFT,
this is a very complicated process and the effective actions at higher
loops are non-local functionals of fields, and of covariant differential
operators and typically with infinite number of terms hence almost
impossible to handle exactly without employing any approximations.
One may think that the gauge symmetries of the theory (if they are
non-anomalous on the quantum level) may help in constraining the number
of terms in the effective action. However, since in EFT, we do not have
to obey the renormalizability conditions at loop orders, the number
of terms is still infinite, and all of them can be written in a gauge covariant
fashion. In the act of resignation, we may contemplate the details
of the procedure of quantization transforming some classical action
functionals into much more complicated non-local functionals at one-loop,
two-loop and higher loop levels. Someone may think about taking some
shortcuts and for example considering two-loop effective action as
a result of quantization applied to one-loop level of already quantized
one-loop level effective action. This is not fully correct and such
one-loop RG-improved actions are valid only approximately. However,
one shall focus on the possibility of an existence of some fixed point
action functional of such procedure of obtaining higher loop corrections.

In addition, here is the place, where the conformal symmetry enters for the
first time in QFT. It happens that, when the conformal symmetry is
present and non-anomalous on the quantum level, then the above procedure
is at the fixed point, or, in more mathematical words, the quantization
procedure is here idempotent and the full effective action (with all
perturbative orders resummed and also with non-perturbative contributions)
is here identical to the classical action of the theory that we started
with. This is a very luxurious situation, when the quantum theory looks
exactly the same as the classical one. Such examples are difficult
to find, but one is already known for a long time. This is an ${\cal N}=4$
super-Yang--Mills theory (SYM), which is very symmetric and moreover
conformal on both classical and quantum level in $d=4$ spacetime
dimensions. The other models result as deformations to the gauge structure
of ${\cal N}=4$ SYM theory and therefore this theory can be viewed
as a ``harmonic oscillator'' of the theoretical physics in the 21st century. All other theories may look more realistic, but their
theoretical description and understanding that we have about them
is only a direct consequence of what we understand about this conformal
theory and a simple afterward application of a conformal perturbation
theory to some operators which deform the original structure. The
explanation why conformal symmetry is essential here for having very
nice form of the effective can be found below.

There is possible a question why theories, which are conformally invariant
on the classical level, cease to be such on the quantum level. One
may blame the fact that the quantization procedure of classical field theories
does not preserve conformal symmetry, but the situation can be in general
more complicated. If the symmetry is not present on the quantum level,
but it was present on the classical one, then we say that the theory
is anomalous. There could be three distinct reasons why the absence of some symmetry may appear at quantum level:
\begin{enumerate}
\item First, the obvious reason is if the classical Lagrangian
of a field theory model does not possess the original symmetry and this is also
inherited by quantum dynamics.
\item  The second reason is related to that, in the definition of quantum theory
via path integral, we have to integrate over field space with some
specific measure. In addition, this measure may lead to anomalies as explained
in the Fujikawa method of deriving, for example, a chiral anomaly.
\item The last reason is that, in the process of defining the quantum theories
and absorbing ubiquitous UV-divergences, we need to specify a renormalization
procedure to give physical meaning to all infinite results of the
theory. This procedure also leads to anomalies, if the scheme of renormalization
is not chosen adequately for the symmetries present in the classical
theory.
\end{enumerate}

In the last two cases, one says that the quantum anomaly shows up. 
Generally speaking, the presence of the quantum anomaly may
be disastrous for the definition of quantum theory, if, for example,
in the proof of renormalizability, this symmetry is used via Ward--Takahashi
or Slavnov--Taylor identities. Therefore, gauge or gravitational (and
also mixed gauge-gravitational) anomalies should be all cancelled
in consistent models of QFT. This is done by careful examining the
spectrum of all particles of the theory, their quantum numbers and
the structure of their interactions. In addition, for example, the standard model
of particle physics (SM) coupled to Einsteinian gravity is anomaly-free
theory but not all of its extensions are. Finally, the requirement
for anomaly cancellation is a strong condition for model building.
Furthermore, we notice that the absence of anomalies is a condition
pertaining to loop orders in expansion of quantum dynamics and, at
tree-level, such requirement does not seem essential. This means that
we can freely and correctly construct all tree-level $n$-point functions
for the theory (which correspond all to classical level) without bothering
about the problems of invariance of the measure in path integral or
a renormalization scheme which could preserve all symmetries. There~the output is simple and classical: if the theory is with symmetry
classically, then all $n$-point Green functions will enjoy this symmetry
and there is no issue of anomaly. On the contrary, on the quantum level,
the anomaly may show up (and does indeed in anomalous theories) and
this can be seen for example by evaluating famous one-loop triangle
diagrams.

What about conformal anomaly? How does this show up in the models
of QFT? The short answer is that it shows in UV-divergences of the
theory \cite{confan,confan2,confan3,confreview}. They are quite common in QFTs so the immediate
conclusion is that many QFTs are with conformal anomalies. However, most
of the time, this is not crucial for features of quantum theories since
physicists got used to dealing with infinities via a renormalization
programme. In addition, since typically for matter (e.g., non-gravitational)
theories on the classical level, conformal symmetry was not gauged
(was not made fully local), then this was not a gauge symmetry of
these models and was not essential on the quantum level in proving
Ward identities, for example. For matter theories, conformal invariance
was some feature of the classical theory generically not enjoyed any
more by the quantum dynamics, which is in turn based on quantum non-anomalous
matter gauge symmetries. The notable exception here is mentioned above
${\cal N}=4$ SYM theory, which is not only quite highly supersymmetric
(eight generators of supersymmetry) but also conformal on both classical
and quantum level. However, here in such superconformal theories on
flat Minkowski spacetime background, the conformal symmetry is not
gauged.

It is quite easy and intuitive to understand why conformal symmetry
has to do with divergences of the theory, when the upper UV limit
of loop integrations is taken to infinity. First, we noted that, in
a very popular dimensional regularization scheme (DR) of renormalization,
it is customary to change the dimensionality of spacetime by a small
amount. However, conformal invariance of the theory is very sensible
to such even small manipulations and a theory which was originally
on the classical level conformally invariant in $d=4$ spacetime dimensions
behaves as not conformally invariant in other dimensions (typically
in $d=4-\varepsilon$). This is related to particular constraints
that the conformal invariance puts on the terms in the Lagrangian and
we write more about this also below. Simply,~if~we do not have divergences,
then in DR scheme, we do not have to leave $d=4$ dimensions and the
theory if conformal remains such also on the quantum level. Another
advantage of using DR is that some infamous UV-divergences which are
not scheme-independent (we mean power-law divergences) do not beset
our renormalization programme in this approach and we have to deal,
and renormalize, and absorb only logarithmic divergences in the cut-off
$\Lambda$ (put as the upper limit of integrations over loop~momenta).

There exists also another argument which links conformal invariance
and the absence of UV-divergences (which is also called UV-finiteness
of the theory). If the theory is truly scale-invariant (which is the
first step towards conformal invariance), then the upper limit of
loop integrals can be freely rescaled so the result of the integration
is independent on $\Lambda$. If it is really independent, then it
means that it is not divergent because the actual value of the integral
can be taken in the case, when~the regulator $\Lambda$ is still finite.
In such conditions, there are no divergent parts of the integrals and
only finite parts are computed in the way described above. The theory
is finite, without divergences, all the beta functions vanish and
the conformal symmetry is present on the quantum level, if the classical
Lagrangian of the theory was conformally invariant too.

Another point of view is to relate divergent parts of the Green functions
computed at loop levels to the divergence of the conformal current.
This gauge current $j_{\mu}$ describes how the theory changes when
the scales of momenta of all particles of the theory are rescaled.
If the theory possesses a conformal anomaly on the quantum level, then
this current is not conserved in the operatorial sense in Hilbert space and
 even in the sense of quantum expectation
values. This has also to do with the trace of the effective energy-momentum
tensor of matter fields considered as a quantum operator. If this
tensor with quantum corrections is trace-free (in the sense of quantum
vacuum expectation values of the corresponding operator), then we
have conformal symmetry present on the quantum level as well.

The last thing that we would like to discuss here is the issue of
conformally invariant regularization of some QFT models. As emphasized
above, if the theory produces UV-divergences at the quantum level, then
it cannot be conformally invariant. However, there do exist conformally
invariant regularizations and conformally invariant forms of the divergent
parts of the effective actions. (They can be obtained by applying
a trick with conformal compensator to any generally covariant terms
in the action.) However, as mentioned above, it is only one of three
needed conditions that is satisfied here. The~three conditions are
needed in order to have full conformal invariance on the quantum level.
The other two were the conformal invariance of the classical action
and of the field measure in the path integral. Only~all three conditions together
secure the conformal symmetry on the quantum level. If the conformally
invariant regularization is used and even if the divergent parts of the effective
action are written in the conformal way, still the theory is not conformal
on the quantum level because still it contains UV-divergences. This
basically means that the integration measure of the path integral
in such models is not preserved by the conformal transformations. We can
give here a simple example of one-loop situation in the ordinary four-dimensional
YM theory. The divergences are there and they can be nicely collected
into a divergent part of the effective action taking the gauge-invariant
form
\begin{equation}
\Gamma_{{\rm div}}=\frac{c}{\varepsilon}\!\int\!d^{4}x\,{\rm tr}F^{2}\,,\label{eq: divym}
\end{equation}
where $c$ is some non-zero constant coefficient depending on the
precise structure of the gauge group and $\varepsilon$ is the regulator
in DR. Actually, here we notice that, due to the renormalizability of YM
theories in $d=4$ spacetime dimensions, this action takes the same
structural form as the classical action of YM theory. We remind readers that,
from the time of Bateman and Cunningham, it is known that in $d=4$
(and~only there) the $F^{2}$ action is classically conformally invariant.
This endorses that also our procedure of renormalization of one-loop
quantum divergences is conformally invariant, since the action $\Gamma_{{\rm div}}$
is. (Here, we do not even need a conformal compensator to achieve this.)
However, everyone agrees that pure YM theory is not conformal on the
quantum level and this can be traced back to the fact that the divergences
are indeed present, even if written in a conformally invariant manner!
The appearance of UV-divergences (regularized in whatever way, even
in the conformal one) shows that conformality is broken and this is
another side of the same coin. If we carefully look at the other side,
we see problems with the conformal invariance of the measure in the
path integral. The absence of UV-divergences or vanishing of the beta
functions (or in the language of the RG community the condition of being
at a fixed point of RG) are all signals that we are dealing with conformally
invariant theories on the quantum level. These all are different sides
of the same coin that can be investigated also from the technical
point of view by the analysis of invariance of the measure, of the
quantum renormalization procedure and of the classical action of a theory.

The first prerequisite for conformal theory on the quantum level is
to have a classically conformally invariant theory. In addition, for this, the
very simple requirement is classical scale-invariance. This means
in turn that, in the classical Lagrangian of the theory, we cannot have
any dimensionful parameters, like masses of fields or dimensionful
coupling constants (like the Fermi coupling of weak interactions).
If~in the theory we have one mass scale, then this is equivalent to
having many of them because we can always rescale it to get any other
energy scale at our wish. We will also comment later that
classical scale invariance is quite restrictive and can determine
the form of the action of a theory to some big extent. For example, the
requirement of the absence of dimensionful couplings in any dimension $d$
first gives rise to renormalizable models. However, the situation with
conformal symmetry on the classical level puts even more constraints and
the renormalizability on the quantum level may not be achieved.
Classical scale-invariance of the theory
requires no mass for fields or any other scales in parameters needed
for the definition of the theory. One may think that the same should
roughly happen on the quantum level, where in the definition of the
theory we may have addition of new scales (related to renormalization
$\mu$). The whole point about quantum conformal theories is to avoid
the introduction of this new scale $\mu$ and then we have exactly
the same situation as on the classical tree-level---no mass parameters
in the definition of the theory. This situation with enhancement of
symmetries on the quantum level is the reason for miracles and simplicity
of quantum CFT.

After discussing at length the issues related to the conformal anomaly
and its effects on the quantum dynamics, now we come back to the situation
with conformal symmetry in quantum gravity. As any symmetry in quantum
gravity, this invariance has to be gauged on the quantum level 
\cite{Kallosh, Banks}. If someone tries to marry conformal symmetry
in gravitation with quantum mechanics, then one encounters very serious
problems with consistency of a such framework. Not only merging of
relativistic gravitation with quantum field theory is difficult, but
on top of problems there, we also have to make the full theory consistent
with local conformal symmetry (invariance with respect to Weyl rescalings
of the metric tensor:
\begin{equation}
g_{\mu\nu}\to g_{\mu\nu}'=\Omega^{2}g_{\mu\nu}\,,
\end{equation}
where $\Omega$ is an arbitrary function of spacetime points $\Omega=\Omega(x)$).
In the language of anomalies, we must ensure that, in a consistent quantum
theory, all gravitational and conformal anomalies are precisely cancelled.
This regarding the conformal part of the full symmetry group (which
will consist here of diffeomorphism and local conformal transformations)
means that, in such theory, on the quantum level, there are no UV-divergences
at any loop order. This is a very strong constraint, almost impossible
to be achieved in models of quantum gravity often considered in the
literature. It is desirable to find a theory, in which all beta functions
(perturbative as well as non-perturbative contributions) are exactly
zero and such theory can be viewed as a gravitational gauged analogue
of the famous and very well behaved on the quantum level ${\cal N}=4$
SYM theory. The fact that all beta functions are tuned to zero is
an expression of very high fine-tuning but also all superconformal
theories are highly fine-tuned. In the quest for quantum conformal
gravity theory, the consistency requirement of the absence of conformal
anomaly is very crucial since this symmetry is gauged. Moreover, it
constrains quite a lot of possible searches for candidate conformal quantum
gravity. This local symmetry is instrumental in proving Ward identities
(here conformal Ward identities) and they are very important elements
in the description of a quantum dynamics of such a system. If we do
not have conformal anomaly of precisely zero, then all features and advantages
of conformal symmetry on the quantum level are inaccessible and lost.

We need conformal invariance preserved at the quantum level of the
theory because this is the level which decides about very important
features of the theory, its renormalizability, control of the divergences
and unitarity. As it is known for particle physics, it is desirable
to have very nice properties of the theory per se, while the specific
choice of the vacuum, asymptotic states or background solutions of
the theory may not respect all of them and may lead to breaking of
some symmetries of the theory. However, for theoretical considerations
about the situation in a theory, it is better to have as many symmetries
as possible since the symmetries constrain the dynamics, lead to
integrability or even to solvability of some models.

In the search for a conformal quantum gravity, we may take various
attitudes and various directions. First, we may desire to analyze the
situation with standard Einsteinian gravity as described by Einstein--Hilbert
action in $d=4$ spacetime dimensions:
\begin{equation}
S_{{\rm E-H}}=\!\int\!d^{4}x\sqrt{|g|}\kappa_{4}R\,.\label{eq: ehaction}
\end{equation}

However, such theory (with or without a cosmological constant term
which could be possibly added here) reveals itself to be non-renormalizable so
the problems here with UV-divergences are even more severe. We write
a detailed account on UV-divergences and their history in this theory
in the section about infinities. Although it is known that in $d=4$ spacetime
dimensions this theory on the classical level is not scale-invariant, because of
the presence of a dimensionful coupling constant $\kappa_4$ related to the
Newton's constant, the situation can be improved a bit by using a trick with a
dilaton field. Such dilatonic Einsteinian gravitation is indeed scale-invariant
classically, but the UV-divergences are inevitable on the quantum level and
they destroy the conformal symmetry. We consider more about the dilaton trick
also in Section \ref{sec:CQG}.

This review is organized in the following way. In the next section,
we review the long story of infinities which at the end led to the
first construction of UV-finite gravitational theory. Then, we comment
on the conformal quantum gravity, its situation at the one-loop level
and also its scattering amplitudes (both at the quantum and classical
levels). Finally, in tight relation to previous topics, we discuss some
 virtues of the Fradkin--Tseytlin conformal gravity models in four dimensions.

\section{Long Story of Infinities\label{sec:Story-of-infinities}}

Interacting QFTs were always plagued by divergences. They appear both
in the UV as well as in the IR limit of integration over loop momenta.
Here, we will consider only infinities showing up in the UV limit of
the theory. They are present in almost any model of QFT, with the
notable exceptions of UV-finite theories, which are convergent in
the UV regime. Original hope was that, in fully consistent QFT, we will
be able to avoid all these divergences and the theory will give us
only finite convergent results. However, soon after the development
of first interacting models of QFT, such an idea was abandoned as too
naive. The problem of infinities motivated and strongly shaped the
research in particle physics and field theory in the last 70 years.
They led to introduction of new research directions of renormalization
and renormalization group flows. The understanding of them forced
us to change our views on the parameters of the physical theory and
on how to distinguish between bare (theoretical and unobservable)
parameters and effective and measurable (physical) parameters. Moreover,
new tools had to be introduced to describe quantum physics, like regularization
methods and beta functions. Actually, a particular area of research
in theoretical physics evolved into a whole new field of quantum RG
flows. These developments also gave a little help to our big quest
of classification of quantum field theories and, thanks to various
problems with UV-divergences, we were able to distinguish two classes
of theories: renormalizable and non-renormalizable as first discussed
by Pauli. Since the first distinction was based on simple analysis
of the energy dimension of couplings, the further progress led to
refined notions of renormalizability and also non-renormalizability.

Before we get to the UV-finite gravitational theories (so theories
completely without divergences), let us first briefly review the situation
with divergences in other theories of gravitational interactions on
the quantum level. These theories were arising due to historical reasons
in the following order. First, gravitational theory based on Einstein--Hilbert
(E--H) action
\begin{equation}
S_{{\rm E-H}}=\!\int\!d^{d}x\sqrt{|g|}R\label{eq: EHaction}
\end{equation}
was quantized and the situation with divergences was analyzed. Since
the conclusions were very pessimistic, this motivated many researchers
to change and modify the classical theory and start with a quantization
of some modified gravity models. The big promise arouse with four-derivative
theories in four-dimensional spacetimes since they were the first models
to be proven to be power-counting and multiplicatively renormalizable.
Indeed, the modifications of gravity, which includes terms with higher
number of derivatives revealed to be fruitful in gaining control over
infinities at the UV limit. For example, gravitational theories with
ten derivatives in four spacetime dimensions revealed to be one-loop
super-renormalizable meaning that the control over divergences was
highly strengthened. In~such models, the divergences appear only at the
one-loop level, while at higher loops the results of all integrals
are convergent. Theories with higher derivatives (HD theories in short)
are typically beset by the problem of ghosts and to solve this issue
non-locality was invoked. In non-local theories, the problem of ghosts
does not show up, but at the same time good control over divergences
can be still kept. Actually, it is possible to construct such non-local
gravitational theories for which the behaviour of the theory in the
UV limit coincides exactly with the behaviour of one-loop super-renormalizable
theories. Then, the final step in the construction of UV-finite theories
is to add some operators called ``killers'' to ``kill'' completely
the remaining beta functions at one-loop level. Actually, we will see
that, in UV-finite theories, the non-locality is not a must and the
theory can solve all the problems of higher derivative formulation,
if it is conformal on the quantum level. However, this is the chronological
path and we will follow such in the presentation here.

First, we consider UV-divergences in E--H theory in $d=4$ four dimensions.
This theory can be quantized in a standard way using the Faddeev--Popov
methods on the path integral. The result of such covariant quantization
is a QFT with one coupling, which is a mass parameter ($M_{{\rm Pl}}$),
which is the square root of the inverse of the gravitational Newton's
constant $G_{N}$ (up to some numerical coefficients). As such, it
is a dimensionful parameter in the theory, which may signal a forthcoming
problem with infinities and renormalizability. This is verified and
confirmed by the careful analysis of perturbation calculus in such
a theory, which is actually a series in $\kappa_{4}=M_{{\rm Pl}}^{-1}$,
which in turn has the inverse mass dimension. According to the criterion
by Pauli, such theory is non-renormalizable (at~least on the perturbative
level) since this series reminds a series one gets if perturbatively
study Fermi's theory of weak interactions. The E--H quantum gravity
has serious problems with divergences---at least this is what the
naive power-counting analysis is telling us. However, maybe we should be
more careful and do not always trust the worst behaviour given by
the power-counting.

Indeed, in 1974, 't Hooft and Veltman were more careful and with a
big effort \cite{hove} computed explicitly the form of divergences
at one-loop in quantum gravity. Their divergent part of the off-shell
effective action takes the form:
\begin{equation}
\Gamma_{{\rm div}}^{(1)}=\frac{1}{8\pi^{2}(4-d)}\!\int\!d^{4}x\sqrt{|g|}\left(\frac{1}{120}R^{2}+\frac{7}{20}R_{\mu\nu}^{2}\right)\,.\label{eq: hoveaction}
\end{equation}

 Their setup was the following. Their theory was quantum Einstein--Hilbert
without matter (pure gravity) treated perturbatively at one-loop in
four dimensions. They also analyzed the situation for both on-shell
and off-shell Green functions of the theory. What they found is that,
contrary to the expectations expressed above, such model was completely
convergent for everything that regarded the on-shell (using classical
vacuum equations of motion of the theory) $n$-point functions. This
is truly a ``miracle'' since, at one-loop, this theory is UV-finite.
(It is not merely a non-renormalizable one, it is one-loop finite!) Thus, 't Hooft and Veltman
achieved finiteness of E--H theory already in 1974. This is not true
because the ``miracle'' does not always happen, especially if we
want to move in any direction away from the setup considered by these
authors. Moreover, the reasons (mathematical and physical) can be
understood for the ``miracle'' to happen and a deeper analysis shows
that, in a more general situation, we should not expect anything like
this to happen again.

Now, thanks to 't Hooft and Veltman, we can explain the result with
a complete absence of divergences at one loop. This is due to two facts.
The first is the possibility of performing gravitational field redefinitions
for on-shell Green functions on the backgrounds, which are gravitational
vacua (that is are Ricci-flat in E--H theory not coupled to any matter).
The second fact is that, in $d=4$, we are lucky and a variation of a celebrated
Gauss--Bonnet term
\begin{equation}
{\rm GB}={\rm Riem}^{2}-4{\rm Ric}^{2}+R^{2}\label{eq: GBterm}
\end{equation}
is a total derivative, and hence can be neglected under spacetime volume
integral (if the proper boundary conditions are used or the fields
of the theory fall off sufficiently fast at asymptotic infinity). We use
the GB density in the action integral of the theory at the one-loop
effective level  and, by variation of the
last object, we derive equations of motion (EOM) and two- and higher $n$-point
Green functions. For~such purposes, if something is a total derivative, then
it can be safely ignored. This last fact has to do really with the number of dimensions because
another way of writing the Gauss--Bonnet term is
\begin{equation}
{\rm GB}=\frac{1}{4}\epsilon_{\mu\nu\rho\sigma}\epsilon_{\kappa\lambda\tau\omega}R^{\mu\nu\kappa\lambda}R^{\rho\sigma\tau\omega}\label{eq: GBepsilons}
\end{equation}
and there we explicitly see the use of the completely anti-symmetric
Levi--Civita tensor $\epsilon$. Such tensor with four indices can
only be defined in four dimensions and in other dimensions we have
a mismatch between the number of indices on two Riemann tensors (eight indices)
and on two epsilon tensors (in~general case $2d$ indices). This explains
why four dimensions are special regarding the properties of the Gauss--Bonnet
term. Altogether, these two facts make the on-shell part of the divergent
action at one-loop level in pure E--H quantum gravity vanishing, so there
is no a problem of divergences~there.

However, the ``miracle'' is gone, if we try to consider more general
setup. If we change the number of dimensions to $d\neq4$ (typically to $d>4$),
then in such new dimensions there is nothing like GB identity and there is
no reason to expect absence of divergences. If we couple matter to
pure E--H gravity, then Ricci-flat manifolds are not background exact solutions
of the theory and we cannot use vacuum field redefinitions any more. Similarly,
if we go off-shell and study amplitudes for virtual external particles,
again we cannot redefine the metric field there. Finally, the situation
needs to be investigated at the two-loop order. First, naive power-counting
analyses suggested that we should expect divergences with energy dimension
6, so a corresponding term in the divergent part of the effective
actions would contain six derivatives on the metric tensor.

Algebraic considerations by Nieuwenhuizen \cite{nieuwe} and others
showed that, in vacuum (Ricci-flat) solutions, all algebraic tensors
that can be constructed with six derivatives reduce to tensors, which
are cubic in gravitational curvatures and moreover, due to assumed
Ricci-flatness, they all can be written as cubes of the Riemann tensor
(various contractions). Actually, the usage of some dimensionally dependent
identities reduces this algebraic space to only one element, which is conveniently
represented using a traceless Weyl tensor. (Since we get the Weyl tensor from Riemann by subtracting some Ricci tensors or scalars,
this change on Ricci-flat backgrounds amounts to nothing.) This one
scalar cubic invariant in Weyl tensor we denote schematically by $C^{3}$
and its explicit form
\begin{equation}
C^{3}=C_{\mu\nu}{}^{\rho\sigma}C_{\rho\sigma}{}^{\kappa\lambda}C_{\kappa\lambda}{}^{\mu\nu}\label{eq: weylcube}
\end{equation}
shows some kind of cyclicity and symmetry of it. (This invariant looks
like a trace of Weyl cube tensor treated as a matrix in its pairs
of indices. Moreover, the contractions of indices occur exclusively
parallely in pairs.) Hence, the question was whether such term could
appear in the on-shell part of the divergent effective action at the two-loop
order in pure quantum gravitation based on E--H action. By a very tedious
computation, Goroff and Sagnotti in 1985 showed that the coefficient
of the counterterm needed to absorb divergences appearing at this
level is non-vanishing \cite{gosa1,gosa2,gosa3}. We just quote the
exact figure for the reference here (but important is only the fact
that it is non-zero):
\begin{equation}
\Gamma_{{\rm 2-loop,\,on-shell}}=\frac{209}{2880(4\pi)^{4}}\frac{1}{4-d}\!\int\!d^{4}x\sqrt{|g|}C^{3}\,,\label{eq: 2loopdiv}
\end{equation}
where this result was obtained in a dimensional regularization (that
is why the factor $d-4$ appears above). This shows that also at two-loop
level the ``miracle'' is gone. This is to be expected since we do
not have any strong reason to exclude UV-divergences of this form
at this level. All facts about field redefinition and speciality of
four dimensions cannot help in reducing algebraically the invariant
$C^{3}$ and, since it is algebraically possible, then it will be generated
by quantum corrections. (This is some kind of a Murphy's law of QFTs:
if some divergent term is not excluded by the symmetry or algebraic
considerations, then sooner or later it will appear in quantum computation.)

In the renormalization of the E--H theory at the two-loop order, we have to be prepared to absorb
any divergence, which appears due to quantum corrections and this
divergence requires quite exotic covariant counterterm (with six derivatives)---compare this one with the original E--H action containing two derivatives.
The non-zero coefficient above signifies that the pure E--H theory
is non-renormalizable at the two-loop order and the naive expectations
from power-counting analysis hold true. To renormalize such theory,
even on-shell, we need a counterterm of the covariant form $C^{3}$
and such was not present in the original action. 

We also note that, from the two-loop level, the expectations based on
schematic power-counting analysis are true and the theory behaves
quite badly regarding the control of UV-infinities. When~we increase
the loop level, the terms arising as divergences have higher powers
of energies, so~we need more derivatives to absorb them in covariant
counterterms. The control over divergences is here completely lost
and the original hope for a consistent QFT of gravity drifts very far
away. We~cannot take E--H action as a basis for quantum field theory
of gravitational interactions. This~theory behaves very badly on the
loop levels, although it can do well as an effective theory (in the
first approximation). Besides the very theoretical problems of non-renormalizability,
the real physical problem is mentioned above non-predictiveness and
this has to do a lot with experiments. \mbox{Actually, a careful} reader can
notice that, due to the renormalization, such theory on higher loop levels
makes the same predictions as a sufficiently complicated theory
with higher derivatives. 

The situation with divergences is very bad in the initial gravitational
theory. What could help us to overcome these problems? First, a bit
conservative solution is to add more symmetries to the theory. This
is fruitful, when symmetries joining fermions and bosons are added
because they are known from playing a crucial role in cancelling divergences
as a result of fine-tuning of interactions between different fields,
but now residing in the same symmetry multiplets. For the case of gravitation,
the addition of supersymmetry (here in local version) leads to supergravitation
theories. These theories on the quantum level indeed behave better
than their just pure gravity counterparts. The situation is improved
a little due to gravitinos and their gravitational interactions. Since
they are fermions, loops with them contribute with minus signs and
this opposite sign reveals to be important for cancellation of some
divergences. However, not all divergences are cancelled. What is worse:
not all divergences are even under control. In original supergravity
models with two or four local supersymmetries, we~know that, at a sufficiently
high number of loops, the divergences show up and they are not possible
to be absorbed in the original two-derivative action, due to the arguments
pretty much the same like elucidated above. Hence, the theory reveals to
be non-renormalizable, not only non-finite, so original hopes are
gone.

In turn, the situation in supergravities with high number of supercharges,
like in ${\cal N}=8$ in $d=4$ is still not settled down, due to the
very complicacy of the theory and it is not known, if~the theory is
with divergences at five loops, for example \cite{smax1,smax2, smax3,smax4}. The quite high level of
loops is, of~course, another tremendous difficulty here. Original and
old conjectures were that such theory may reveal to be UV-finite;
however, if the divergences are present, then it will be non-renormalizable
like any other (super-) gravity theory in $d=4$ based on two-derivative
action in pure graviton sector. Indeed, this is true, that all supergravities
theories discussed so far in this paper in the gravitons sector coincide with a
theory based on E--H action, since they all have two derivatives. The
logic as laid down above will apply to them too, since in their actions they have two
derivatives only, and in four dimensions this is too little. We think
that right now it is likely that even very highly constrained by local
supersymmetries (and other hidden symmetries) models of supergravity
with two-derivative actions will show at sufficiently high level of loops
divergences, which cannot be absorbed in the existing terms of two-derivative
theory. Actually, the problems there are formulated in a form of questions,
whether the counterterms can exist or not at all, due to the all existing and constraining symmetries.
The miracles at one- and two-loop levels are due to the facts that
some counterterms (like~$C^{3}$) in supergravity cannot be written
in a (super-)symmetric way, so they cannot show up and in the result
there are no divergences coming with them. However, the superspace
formulation does not exist for theories with ${\cal N}>1$ supersymmetries,
so it is also very difficult to use such covariant formal arguments.
We leave the path of highly constrained two-derivative supergravity
theories, since now they do not offer very convincing and satisfactory answer for
the problems of UV-divergences.

Another option is to try to employ conformal symmetry already on the
classical level in gravitational theory. However, since E--H is a theory
with only diffeomorphism symmetry which is gauged, this is difficult.
Some attempts can be done with a dilaton field (treated like a dynamical
conformal compensator), but generally the situation with infinities
is not improved. In addition, this can be traced back to the fact that the
gravitational coupling constant has dimension of mass or equivalently
the dilaton field takes a vacuum expectation value (v.e.v.). If we
have a theory with a mass parameter, it cannot be on the classical
level scale-invariant. Strictly speaking, here we have a vacuum state
of the theory, which is not scale-invariant, but the full theory is
scale-invariant. Hence, we do not start with a setup for a conformal
gravity theory on the classical level. If we manage to have such a
theory as the starting point, then another pending question is whether
it remains conformal also on the quantum level. This question is even
more important than the original one, since we have examples of classically
conformally invariant theories, which are not any more scale- and
conformally invariant on the quantum level, like the standard two-derivative
YM theories in four dimensions. 

There could be also a possibility that
a theory, which is not classically conformally invariant, becomes
such after inclusion of all quantum corrections, but, until recently, we have not
known any example of such a behaviour. Such quantum enhancement of 
symmetries actually appears in the quantum gravitational setup
that we are going to consider in this paper. The examples of such enhancement 
of symmetries on the quantum level are already known in the theory of critical phenomena. 
There, due to quantum RG effects, the special point at  sufficiently high energies is met. 
The symmetry of the theory at this conformal fixed point of RG flow is bigger than the original classical theory at small energies. The quantum effective action
at this critical point can be viewed as a resummation 
of perturbative quantum contributions with all effects of RG taken into account. 
This effect can be clearly seen as the non-perturbative one (or as we want to phrase it as
 arising from resummation). The situation with the theories 
of quantum gravity that we analyze in this review is such 
that already at the one-loop level all contributions of perturbative 
loops are summed up (at least regarding UV-divergences). 
This amazing feature is due to the super-renormalizability 
property of higher derivatives theories that we are concerned with in this paper.
The quantum enhancement of symmetries could be also realized, if
the dilaton field dynamically decouples and drops out from the effective
action of the theory. In a sense, this would be the reverse of a dimensional
transmutation process, which in some theories without scales produces
new scale, which arises dynamically, like the famous $\Lambda_{{\rm QCD}}$
scale arising in the theory of strong interactions.

It is clear from the preceding discussion that we must modify the
gravitational theory on the quantum level, and by this we do not mean
adding more symmetries, but change completely the underlying dynamical
principles governing quantum gravitational field. Below, we stipulate conditions that we want to preserve,
when putting modifications to gravitational theory on the quantum
level. First, we want to use and accept standard rules of Quantum Mechanics
(QM) and, as a consequence, we will modify gravity part, keeping the quantum mechanical part untouched. Following these rules, we want to cast
the quantum theory of the gravitational field in a framework of very
successful (so far) QFT. To quantize a classical system with gauge
(here diffeomorphism) symmetries, we will use standard covariant rules
of quantization of gauge theories. We can use, for example, a path~integral
approach to quantize the theory and to obtain generating functionals
for Green functions. For~dealing with gauge symmetries, we will use
a standard Faddeev--Popov trick of quantization of gauge symmetries.
In this way, we hope to get a standard version of QFT of gravitational
interactions. As marked above, we will use covariant treatment, which
manifestly preserves Lorentz symmetries and, moreover, is~background-independent---this is why we will use a formalism of background field theory.
Additionally, we want to exploit the power of perturbative framework
of QFTs. At the end, what we want to get is a consistent QFT framework
for gravitational interactions.

From the point of view of particle physics, we propose a QFT, whose
particle excitations correspond to structureless and point-like massless
gravitons with two helicities and with the spin 2. Setting the rules
above, now we take the step to modify classical gravitational theory,
which is later quantized according to the stipulated rules. We modify
the gravitational theory by changing its action, since the action defines
completely the dynamics for field theory systems. We decide to take
this step as an act of desperation, since all other attempts failed
to give us satisfactory answers for the problem of UV-divergences in quantum
gravity theory. We know that, in the EFT framework, Einsteinian gravity
 is a good physical theory, but we are also aware of its limited range of applicability.
It works very well on the classical macroscopic level (like the distances within
our solar system). However, it does not describe the dynamics very
well at the very small scales (microworld) and at very large scale (cosmology).
Therefore, it is not surprising that we must modify this theory at
those scales and that a more general theory should be used for quantization
programme, while the Einsteinian gravity should be recovered only
in the classical macroscopic limit and for low energies.

We desire the following modifications done to the gravitational theory.
We modify the gravitational action of a theory in order to satisfy
many different requirements (some of them really sound like perfect
wishes). Namely, we want a self-consistent theory of massless spin-2
interactions on the quantum level, so the theory must be written using
covariant language, using gravitational curvatures and covariant derivatives only.
Such theory should be defined at any energy scale, so we should not
find any problems, neither in UV, nor in the IR limit of the theory.
This theory should be possible to be formulated in four-dimensional
spacetime. This does not preclude that we will not accept an idea
of extra dimension(s). We just want to say that the theory should be
first defined in four dimensions with the possibility for extension,
it should not be only constrained to $d=4$. 

As for the construction of a gravitational theory, we want to use only
metric degrees of freedom and we will not allow for dynamical torsion
or non-metricity. However, in the gravitational spectrum of the theory,
we may want to accept spin-0 and spin-1 particles as well, together
with a dominant massless spin-2 contribution from gravitons. Regarding
the strictly quantum properties, we want a theory written in a QFT
language, which will lead to unitary $S$-matrix for scattering processes
between on-shell asymptotic states of the theory. Since Lorentz-invariance
is well tested experimentally and holds true to very high energies,
we also want to preserve this symmetry of Nature in our construction.
Lorentz or more generally Poincar\'e{} symmetry is a symmetry of
asymptotically flat spacetimes, hence this symmetry pertains to the
definition of asymptotic states, which are used for the construction of
the $S$-matrix. We do not want to break this good and useful symmetry
either. Moreover, in the same spirit, but on general backgrounds,
we do not want to single out any particular geometrical structure, hence
in the middle of our manifold (in not asymptotic regions)
we have to be able to write our theory
in a generally covariant and background-independent fashion. As motivated
extensively above, we must write a new gravitational theory using terms
with higher derivatives added to the E--H action. We will also soon give
another arguments, why higher derivatives are inevitable in quantum
gravitational theory. In addition, finally we express our desire that the situation
of this theory
in the UV is much more improved compared to the original E--H theory.
We want to build a theory, which is UV-complete in the high energy
regime, where the full control over UV-divergences is gained and which
is renormalizable, super-renormalizable, and finally the most difficult
wish is to require from such theory to be UV-finite. In the next pages,
we will show that it is indeed possible to increase the control of
infinities and eliminate them completely in a sequence of steps as
mentioned above.

The final argument, which will be used for the inclusion of terms with
higher derivatives in the gravitational action, has to do with maybe
less ambitious situations than fully-fledged quantum gravity.
Before constructing such theory
of quantum gravity on a general background, we should maybe study
the situation of quantum matter fields on some fixed spacetime geometry,
which may not be flat, but~curved due to the gravitational effects.
For the moment, we can freeze the gravitation and consider it as a non-dynamical
external field present in the system of QFT of matter fields. The~theory of quantum fields on curved backgrounds is well studied and
well understood. This is that less ambitious idea: let us first quantize
matter and put it on a classical background, and for the moment do
not attempt to quantize gravity. This situation has its physical importance
and significance as well, since all the quantum matter fields have
to be treated like, for example, when they are studied on cosmological
backgrounds. The Friedmann--Robertson--Walker (FRW) classical background
spacetime is a very good approximation to gravitational physics happening
some time after the cosmological
Planck era. There~with good accuracy we can describe effects of the existing
gravitational field by putting the quantum theory on curved background.
To give an example, we can mention that to get the correct results
for big bang nucleosynthesis, which produces all known chemical elements,
we~must consider effects of cosmological gravitational field, but it suffices that
this field be considered completely classical. This approach and motivations first
appeared in works on gravitational field by Utiyama and were also repeated
recently by Shapiro.

What can quantum field theory of matter fields on curved backgrounds
teach us regarding quantum gravity? We may consider the issue with UV-divergences there,
when we do not have gravitons on internal lines of Feynman diagrams,
but we can have them on external lines emanating from the diagrams.
Such semi-classical approach to quantum gravity signifies that we
are in a fixed classical external background (external graviton legs
correspond to classical interactions with smooth external geometry),
but at the same time we do not allow gravitons to propagate (not yet)
and therefore there are no virtual gravitons in any part of the diagrams.
The moral of QFT on curved background is that gravitationally interacting matter fields
will generate divergences, for which absorption we will need purely
gravitational counterterms constructed with the help of the background
external metric field. Of course, these divergences are besides UV-divergences
to be found in matter sector, which~are roughly the same as the ones
we could find in a matter sector, when this one is considered on flat
Minkowski spacetime background. (The UV-divergences of the theory
probe the coincident limit of two points on the manifold, and the spacetime
and internal space too are effectively flat for them in this limit.)
Thus, let us analyze these new divergences, which have to be absorbed
by generally covariant expressions constructed with the background
metric. Let us consider a situation, in which the original matter theory
is a renormalizable two-derivative theory and without any dimensionful
coupling, if it was studied around flat spacetime. For simplicity,
we will below focus on the situation at the one-loop level, but the same can
be generalized to any loop order. For the case of one-loop off-shell
divergences, we find using the formalism of QFT on curved background
that the ``gravitational'' divergences are proportional to the Lagrangian
densities:
\begin{equation}
\sqrt{|g|}R^{2}\quad{\rm and}\quad\sqrt{|g|}C^{2}\,,\label{eq: ct4dim}
\end{equation}
where $R$ denotes the Ricci scalar and $C$ the Weyl tensor. We have
the explicit expression for the second counterterm in four spacetime
dimensions (using other more common curvature tensors):
\begin{equation}
C^{2}=R_{\mu\nu\rho\sigma}R^{\mu\nu\rho\sigma}-2R_{\mu\nu}R^{\mu\nu}+\frac{1}{3}R^{2}\,.\label{eq: weyldef}
\end{equation}

Of course, these two counterterms are non-vanishing only on non-trivial
gravitational backgrounds and that is why in flat spacetime theory
of matter sector we never had to worry about them. They, however, arise
on a curved background (like this one given by the cosmological setup earlier,
but there we have $C=0$ on FRW). These are divergences, which should be absorbed
by the terms constructed with the external metric, and we see
explicitly that they contain four derivatives of the external metric tensor.
This is easy to understand since, if the matter theory is without any
dimensionful coupling and it is renormalizable in $d=4$ spacetime
dimensions, then these ``gravitational'' divergences must come with
dimensionless coefficients and this means that they must multiply
terms, which are constructed with precisely $d$ derivatives of the
dimensionless metric tensor. In addition, such tensors in four dimensions are
precisely $R^{2}$ and $C^{2}$ (when, for the moment, we neglect the
third one: Gauss--Bonnet term (GB) and the fourth one: $\square R$,
which is a total derivative). Only in this way can we construct necessary
counterterms to absorb divergences, which generically show up there
in matter theory. We~conclude here that, in general, if we are in $d$
dimensions, the divergences generated by the matter sector on a curved
background contain exactly $d$ metric derivatives, even if the original action
that we would like to use to study propagation of gravitational field
contains only two derivatives (and then it is a simple generalization
of the Einstein--Hilbert action in general dimension). To absorb these divergences, we
need gravitational counterterms with precisely $d$ derivatives. We notice
also that these new divergences with $d$ derivatives are induced due to the
quantum matter corrections, even if the matter theory contains only two
derivatives (or generally less than $d$ derivatives).

Let us recapitulate how to construct higher derivative theories in general dimensions ($d>2$).
The~reason why the physical theories have to be higher derivative (HD) theories is simple:
the observed dimensionality of spacetime is not 2. If this was exactly the case seen for
macroscopic spacetimes, then two-derivative E--H theory would be enough (and such theory is
renormalizable in $d=2$ dimensions). Instead, we see higher dimensionality of
physical spacetimes and with the above motivations this leads us to considerations
of HD theories. The proper HD theories in $d$ dimensions must be constructed from
terms with precisely $d$ derivatives on the metric tensor. For this, we can use
various curvature tensors (Riemann, Ricci tensors or Ricci scalars) and we can
act with covariant derivatives on them. The construction is not restricted only to
the $\frac{d}{2}$ powers of the Weyl tensors, like it was above for the on-shell
divergent parts of the effective action at higher loop orders in the non-renormalizable
E--H theory. In the result, we get some new actions with some new HD terms.
Moreover, the coupling parameters in front of these terms are all dimensionless
in a general dimension $d$.

The first example of a higher derivative gravitational theory was the
theory proposed in 1977 by Stelle \cite{stelle1,stelle2}. Its action
(in four dimensions) reads:
\begin{equation}
S_{{\rm grav}}=\!\int\!d^{4}x\sqrt{|g|}\left(\kappa_{4}^{-2}R+\alpha_{R^{2}}R^{2}+\alpha_{C^{2}}C^{2}\right)\,,\label{eq: stelle}
\end{equation}
where the coefficients $\alpha_{R^{2}}$ and $\alpha_{C^{2}}$ are
dimensionless. This last fact is peculiar to spacetime of dimensionality
four, but the structure of the fundamental four-derivative action
in general dimension is only modified by one additional term according to
\begin{equation}
S_{{\rm grav},\,d}=\!\int\!d^{d}x\sqrt{|g|}\left(\kappa_{d}^{-2}R+\alpha_{R^{2}}R^{2}+\alpha_{C^{2}}C^{2}+\alpha_{{\rm GB}}{\rm GB}\right)\,,\label{eq: quadgrav}
\end{equation}
where GB denotes the Gauss--Bonnet term, and again $\alpha_{{\rm GB}}$
has the same energy dimensions as $\alpha_{R^{2}}$ and $\alpha_{C^{2}}$.
In four dimensions, the last added term contributes nothing to the
dynamics since its variation \mbox{(to any order)} is a total derivative,
hence it has no impact on the equations of motion (EOM), nor on the propagator,
etc. In higher dimensions, this term contributes to the variations
(for example in Lovelock gravitational theories); however, it still contributes nothing to the propagator of graviton around flat spacetime
in any dimension. This theory is a first example of a multiplicatively
renormalizable gravitational theory in $d=4$ spacetime dimensions.
This is an effect due to an improved behaviour of the propagator, which
now falls in the UV limit not like $k^{-2}$ (which was the case for the
E--H theory with two derivatives), but like $k^{-4}$ and this is really
the consequence of having more derivatives in the gravitational action.
The increased power exponent on the propagator is essential for having
renormalizability because now the energy scaling in the diagram of
every internal line is decreased, so~we have lower bounds on the superficial
degree of divergence of any graph (more about this also later). However, we remark
here that this theory is only renormalizable in four dimensions because
only there the coefficients $\alpha_{R^{2}}$, $\alpha_{C^{2}}$ and
$\alpha_{{\rm GB}}$ are dimensionless. (In higher dimensions, the
first renormalizable theory must contain terms with more derivatives.)
Due to renormalizability, the UV behaviour of Green functions is improved,
but still there are infinities and the theory is merely renormalizable
(similar like $F^{2}$ gauge theory in $d=4$ is renormalizable),
but our goal is to find a candidate for UV-finite gravitational theory
(similar to ${\cal N}=4$ SYM theory in $d=4$). This is a step in
a good direction and with it we write gravitational theory in a very
similar way to the theories of other interactions present in the SM
of particle physics. Now, both the matter and gravitational theory
are put on the same footing as theories quadratic in corresponding
field strengths or curvatures. We add that the term $\kappa_{d}^{-2}R$
in the action Eqn. \eqref{eq: quadgrav} was left there in purpose, although it
contains only one power of the gravitational curvature. We keep it
for the correspondence with Einstein's gravity in the low energy
(that is in the IR) limit of the theory. This is not only a limiting theory,
when the derivatives of the metric tensor are small in some sense,
but also when the gravitational curvatures are small. We could also include
a cosmological constant term $\lambda$ in the action, which contains
no derivatives and no curvatures, and hence it would be the first one we
have to write in the EFT paradigm for gravitational theories.
For completeness, in a theory in $d$ dimensions, we should include,
in principle, all terms with up to $d$ derivatives of the metric,
if they are not excluded by any symmetry of the theory. They are needed for good
IR limits of HD theories. Einstein--Hilbert and the cosmological term are
therefore often necessary for a HD gravitational theories in $d=4$.

However, with the advantages of four-derivative theories, we are also meeting
some drawbacks here. On the classical level, there are new solutions
compared to solutions in standard two-derivative theory. The closer look at them
shows that they are typically runaway solutions, hence quite unwanted
(but compare with an eternal cosmological inflation, which is actually
a runaway solution in $R+R^{2}$ Starobinsky theory). On the other side,
we remark that in the first four-dimensional renormalizable theory Eqn. \eqref{eq: stelle},
all Ricci-flat spacetimes are also exact vacuum solutions, similarly to
the situation of vacuum in the E--H theory. The presence
of higher derivatives signifies also the so-called Ostrogradsky instabilities
of the theory. This has to do with instabilities of the classical
exact solutions of the theory, when written in the Hamiltonian formalism.

The negative sign of the ``massive''
term testifies to the negative energy of the corresponding state (due
to these negative energies these fields are also called phantoms).
Of course, we can reformulate the theory (to change the vectors of state)
to provide positive energy for the ghosts, but then the massive pole
will have a negative residue and therefore the corresponding norm
in the Fock space of states will be negative. Hence, the massive spin-2
particles have negative energy or a negative norm and are therefore
non-physical. They cannot appear on external on-shell legs of any
diagram. The negative norm states cannot be removed from the physical
spectrum and hence unitarity in HD theory is violated. Unlike the
Faddeev--Popov (FP) ghost, which cancels with non-physical components of
the gauge fields and hence preserves unitarity, the appearance of
the massive spin-two ghosts contradicts unitarity. Note that the problem
of unitarity is the most difficult in higher-derivative gravity. We
noticed that in such model we find a presence of massive ghost states
with a negative sign of the residuum in the spectrum. This can be interpreted
in another way that the putative state of the gravitational vacuum
is very unstable and immediately decays (by quantum corrections this
is not immediate, but still very rapid decay) and massive ghosts with
negative energies are instead produced. Therefore, this argument we
may take as a sign that the putative vacuum that we use in the definition
of Fock space for HD theories is not the true gravitational vacuum of
the theory and we should look for another vacuum state, which is stable. These
are all sides of the same coin, which is the violation of unitarity
in such class of theories. Lack of unitarity is the general reason
why we do not consider this higher-derivative gravity as the first and
 the best candidate
for the quantum gravity description. At~the same time, this theory
is a very attractive ``toy model'' for the ``true and
fundamental quantum theory of gravity''. Of course, the
problem with unitarity appears in almost any higher derivative theory, whether in gravitational, gauge, or scalar sector.

\textls[-15]{It is required that we comment here on the interesting situation regarding the renormalizability
in the Weyl square gravity. In $d=4$, the theory is non-renormalizable from two loops \mbox{on \cite{FT1, FT2, FradkinT3}}. However, at the one-loop level and exploiting conformal background gauge, the theory is renormalizable.}

We mentioned that in the search for UV-finite theory the next step
is a construction of a super-renormalizable theory. We simply get there
by considering the following generalization of the four-derivative Stelle
theory (where we understand $\square=g^{\mu\nu}\nabla_\mu\nabla_\nu$):
\begin{equation}
S_{{\rm grav}}=\!\int\!d^{4}x\sqrt{|g|}\left(\lambda+\kappa_{4}^{-2}R+\sum_{n=0}^{N}\alpha_{R,n}R\square^{n}R+\sum_{n=0}^{N}\alpha_{C,n}C\square^{n}C\right)\,.\label{eq: hihider}
\end{equation}

(In principle, we could also add here the higher order generalization of the Gauss--Bonnet term: 
\begin{equation}
{\rm GB}_{n}=R_{\mu\nu\rho\sigma}\square^{n}R^{\mu\nu\rho\sigma}-4R_{\mu\nu}\square^{n}R^{\mu\nu}+R\square^{n}R\,,\label{eq: genGB}
\end{equation}
but we notice right away here that it does not contribute to the propagator
of graviton around flat spacetime in any dimension, so it will not
have any impact on the (super-) renormalizability properties.) Such
theory was proposed for the first time and studied by Asorey, Lopez and Shapiro
in 1996 \cite{asorey}. The~feature of its super-renormalizability
we discuss in some details below. Here, we would like to say that this
theory enjoys even higher suppression of the graviton propagator in
the UV, so the loop integrals are made even faster convergent, hence
this super-renormalizability property is not unexpected here. We~will
see that, for a sufficiently high natural exponent $N$, the UV-divergences only 
remain at the one-loop level. In such theory, the graviton propagator (and
also for all other fields in the gravitational sector, like FP ghosts)
scales like $k^{-(2n + 4)}$ in the UV regime. Again here, we find
the same drawbacks of higher derivative theories,
related to ``bad'' ghosts on the quantum level (not to confuse them with ``good'' FP ghosts).
Now, due to the number of derivatives
higher than four in the action, we have the situation that the sign
of the mass terms for ghosts alternate and we have some set of massive
spin-2 normal particles and massive spin-2 ghostly particles, respectively.
Now, we can also understand the meaning of monomial or polynomial behaviour
of the theory, since the sums in Eqn. \eqref{eq: hihider} define two general polynomials
in the box (that is a covariant d'Alembertian) variable. The comments
about the structure of the perturbative spectrum around flat spacetime
apply here as well. The problem of unitarity is still present here and it is not alleviated
by using gravitational actions with higher than four derivatives, for any fixed natural $N$.

The unitarity, instead, does not depend on the question of dimension,
 provided that we are in $d>3$ (to have all the characteristics of
 the relativistic gravitational field turned on). In principle, one could tell
 about the unitarity by analyzing the perturbative spectrum of the
 theory around any background. However, the more refined analysis, which
 also counts the number of degrees of freedom and their linear stability
 requires the usage of Hamiltonian formulation of the model \cite{Kluson, Riegert}.
  Therefore, the latter one can be viewed as a non-perturbative and a fully nonlinear analysis.
  In Weyl gravity, such perturbative analysis points towards non-unitarity; however, some non-perturbative results
  seem to support opposite conclusions \cite{T1, T2, Kaku, Bender1, Bender2, Tkach, Smilga}, which show that either ghosts do not show up or that they are rather innocent, if present. 
 
The final resolution of the problem with unitarity is given by the next
upgrade of theories to fully non-local, so to theories containing an
infinite number of derivatives ($N=\infty$ schematically). All~terms
in the action of such theories could be collected in a form
of non-local form-factors, which can be viewed simply as an analytic
functions of the covariant differential operator, for example of the
covariant box operator ($\square=g^{\mu\nu}\nabla_{\mu}\nabla_{\nu}$).
Such theories are called weakly non-local because, in their presentation
in the action of a theory, all fields and derivatives in the Lagrangians are taken in the same
spacetime point. This is to distinguish them from truly non-local (strongly
non-local) theories for which descriptions we would have to use kernel operators
and the fields depending on different points of spacetime under the
same spacetime volume integral. To handle the latter types of
theories is more difficult but possible as well. However, the non-local
gravitational theories presented in a weakly non-local form suffice
to solve the problem with massive unwanted ghost states. The first
non-local theories were constructed by Efimov, Kuzmin, Tomboulis and
Krasnikov. For the purpose of studying the most general propagator
around flat spacetime, it is enough for us to study the following general
theory, given by the following gravitational action:
\begin{equation}
S_{{\rm grav}}=\!\int\!d^{4}x\sqrt{|g|}\left(\lambda+\kappa_{4}^{-2}R+RF_{R}(\square)R+CF_{C}(\square)C\right)\,,\label{eq: twoff}
\end{equation}
where there are present two form-factors of the covariant box operator
$\square$, namely $F_{R}(\square)$ and $F_{C}(\square)$. (Again,
we can here neglect the non-local Gauss--Bonnet term, which contributes
nothing to the propagator around flat spacetime.) We want to emphasize
that, for the issue of power-counting of UV-divergences, which we
are going to discuss soon, we can restrict
ourselves to the case of flat Minkowski spacetime background because
UV-divergences are independent of the background used to evaluate
them, due to the background independence of the theory. In the UV limit,
any manifold, which is a background, is viewed effectively as a flat
one (UV limit is a coincidence limit of two points). Therefore, the
power-counting analysis we will perform on flat spacetime and for
this purpose we need the most general expression for the gravitational
propagator there. Such expression one can easily obtain from the action
above, which is indeed the most general one on the flat spacetime.

The action above Eqn. \eqref{eq: twoff} is an action of a weakly non-local theory,
 due to the presence of
two non-local functions of the box operator. In general, they should
be analytic functions of the box operator, but in principle the case
of polynomials is also covered here. Moreover, as proven by Tomboulis
\cite{tomb1,tomb2,tomb3}, there exist analytic functions such that
around real axis on the complex plane they tend in the UV limit (regime,
where the argument of the form-factor is real and takes large values)
to polynomials. These~polynomials we would call UV polynomials of
the form-factors. This is a very important discovery in mathematical
physics, since it gives rise to theories that can completely avoid
the problem of ghosts in the perturbative spectrum, but at the same
time take all advantages of the HD theories regarding the control
over UV-divergences. Indeed, as can be easily checked on particular
examples in such non-local theories, we can have no massive ghosts in
the spectrum. The structure of the form-factors is due to this being a bit
constrained, but they can be still quite general and in a large class
of non-local theories. If we use form-factors due to Tomboulis, we
indeed avoid the problems with ghosts, but at the UV limit we get back to HD theories
described by polynomial functions, not by general analytic functions.
The poles of the propagator giving rise to unwanted massive ghosts
would show up, if the expression for polynomial HD theories would be
valid at all energy scales. However, this polynomial behaviour is valid only in the UV limit.
At all ranges of energies, the exact expression is given by the non-local
form-factor, which does not develop any pole on the complex plane, besides
the massless spin-2 pole of the standard transverse graviton. Hence, these dangerous
poles of the HD limit of the theory are completely avoided, due to
the usage of interpolating form-factors. In this way, the unitarity of
the theory is explicitly preserved and there is no any problem with
violation of it, which was present in almost any truncation of the non-local
theory to local HD theories. We saw that in HD theories the occurrence
of unwanted massive ghost states is often inevitable. However, in non-local
theories, we can easily avoid them. Using non-local theories, which
in the UV take back the form of HD theories, we omit ghosts and at
the same time we can still have a very good situation regarding UV-divergences.
A form-factor, which in the UV regime takes the form of any HD theory
is always possible. We emphasize here that, about the UV-divergences,
only the UV physics (so the UV limit of the theory) decide, so we
could completely separate the two issues: of the renormalizability (control
over UV-divergences) and of the unitarity (presence of massive ghosts).
That thing was not possible for HD theories and this is the real advantage
of non-local gravitational theories. As it is well known, the issue of
the spectrum and the presence of massive ghost modes in it is an issue
for all energy scales, not only in the UV physics. That is why it was possible
to disentangle these two problems of renormalizability and unitarity.
In the class of non-local theories, we are able to solve both of them
at the same time. We~add that, in these theories, not only the spectrum
is ghost-free, but also the optical theorem is satisfied and unitarity
of the $S$-matrix is proven rigorously to all loop orders too \cite{unita,unita2,fakeunit}.
In the perturbative spectrum of the theory, we find only physical transverse
massless graviton with two helicities and this is true for both the propagator
around flat spacetime as well as around any maximally symmetric spacetimes
(MSS). The detailed exposition of non-local theories can be found
in the papers \cite{superrenfin,universality,confgr,exactsol,spcompl,RGsuperren,entanglement,evaporation,evaporationconf,fingauge,finmss,finproc,mssfin,review,scattering,singproc,starobinsky,effaction,effactionproc}.

In the theories, which in the UV limit are described by some HD theory,
we observe the following interesting behaviours in the UV regime. First,
the theory is asymptotically free in $d=4$ dimensions similarly to
the situation with Yang--Mills theories, therefore this opens a possibility
for a unification of gravitational and gauge interactions, since they both
have very similar UV behaviour. In such case of HD theories, we also
have very good fall-off of the propagator in the UV limit, very good
control over UV-divergences and we are on a good path towards UV-finite
theories. The quantum loops simply behave very well and in a controllable
way.

Now, let us consider in more detail the situation with power-counting
of UV-divergences in a theory, which in the UV limit takes the following
schematic form:
\begin{equation}
S_{{\rm UV}}=\!\int\!d^{d}x\sqrt{|g|}{\cal R}\square^{\gamma}{\cal R}\,,\label{eq: uvlimit}
\end{equation}
where by ${\cal R}$ we denote a general gravitational curvature (this
could be Ricci scalar, Ricci tensor, Riemann tensor or even a Weyl tensor).
We noticed that with this choice in the UV the theory has a monomial
asymptotics with the exponent $\gamma+2$ on the box in Fourier space
as the kinetic operator between gravitational fluctuations (however, $\gamma$
can be here any positive real number; it does not have to be an integer).
If the theory has as UV behaviour the general polynomial or even a
finite Puiseux series form, then still for the superficial and the
worst behaviour of UV-divergences we can take the situation, which is
given by the highest monomial (with the biggest exponent) and this
exponent is what we above called $\gamma$. Such circumstances we can easily
achieve, if all the two form-factors in  Eqn. (\ref{eq: twoff}) have the
same UV asymptotics. Then, the propagator of all modes of the theory
(of gravitons and also of FP ghosts) has the same scaling in the UV,
as the first time showed by Modesto \cite{modesto}, and this is given by
\begin{equation}
\Pi\sim k^{-(4+2\gamma)}\,.\label{eq: prop}
\end{equation}

In the analysis below, we assume that the graviton field $h_{\mu\nu}$
and a Faddeev--Popov ghost field $C_{\mu}$ (and $\overline{C}_{\mu}$
for a FP anti-ghost field) are dimensionless and this signifies that
for these two propagating fields of the theory we will have the same
maximal number of derivatives in vertices as well as in propagators
of the theory in the UV limit. This also shows that the power-counting
analysis of divergences for the case of quantum gravity in any dimension
is much easier than the same analysis attempted at a higher derivative
gauge or matter theories. As a direct consequence of what was said
above, we see that the maximal scaling in the UV of any vertex of the
gravitational theory (which may also contain precisely one FP ghost and one anti-ghost
field) is
\begin{equation}
V\sim k^{4 + 2\gamma}\,,\label{eq: vertex}
\end{equation}
which is exactly the inverse of the scaling of the propagator $\Pi$
from the formula Eqn. (\ref{eq: prop}) above.

Collecting all above facts, we derive the general expression for the
superficial degree of divergence $\Delta(\mathfrak{G})$ of any Feynman
graph $\mathfrak{G}$ in the theory. Let us imagine that such a diagram
contains precisely $L$ loops (so $L$ integrations over momenta are
to be done), $V$ vertices (of any type: gravitational with any integer
number of legs bigger than two, or also vertices with precisely two
FP ghost fields and any number of graviton legs attached) and precisely $I$
internal lines (which correspond to virtual gravitons or FP ghosts).
Actually, here we can simplify discussion considerably by considering
only gravitons, since, for all that regards FP ghosts, they are very
similar to virtual gravitons. Their energy dimensions are the same
and they have the same UV scaling of the propagator and the same scaling
of any vertex. Let us then, from now on, write only about gravitons,
since adding FP ghosts to this analysis will not change anything.
For any graph $\mathfrak{G}$, we have the following expression for
$\Delta(\mathfrak{G})$:
\begin{equation}
\Delta=dL+V[{\rm vertex}]+I[{\rm propagators}]\,.\label{eq: superficial}
\end{equation}

Since, as emphasized above, we have
\begin{equation}
[{\rm vertex}]=-[{\rm propagator}]=4+2\gamma\label{eq: vertandprop}
\end{equation}
and taking also the topological relation for any diagram ($I-V=L-1$)
into account, we can write the final bound for the superficial degree
of divergence, derived in any dimension $d$, as
\begin{equation}
\Delta\leqslant d-(2\gamma+4-d)(L-1)\,.\label{eq: bound}
\end{equation}

Now, let us particularize to the case of $d=4$. There, we get the following
expression for the bound on the superficial degree:
\begin{equation}
\Delta(\mathfrak{G})\leqslant4-2\gamma(L-1)\,.\label{eq: bound4d}
\end{equation}

Some immediate observations based on this power-counting are here
in order. When we tend to increase the exponent $\gamma$ on the box
in the action Eqn. (\ref{eq: twoff}), then we see that the UV behaviour
given by $\Delta(\mathfrak{G})$ is better and better regarding higher
loops with $L>1$. We see that, for some sufficiently high value of
the parameter $\gamma$, we have the degree $\Delta(\mathfrak{G})$, which is negative and this
signifies by the token of power-counting analysis in the cut-off $\Lambda$
that there are no divergences in the theory at the given loop
order. However, we see that always at the one-loop level we find divergences,
no matter how high we increase the exponent $\gamma$. Moreover, we
also noticed here that at the one-loop level the superficial degree
(so the worst situation) of divergences is equal to the dimensionality
$d$ of the spacetime, at least when we use the cut-off regularization
scheme. Being in $d=4$ spacetime dimensions, we find that, for $\gamma\geqslant3$,
there are no divergences at two loops and onwards. In this situation,
we say that the theory is one-loop super-renormalizable, since the
one-loop level is the last one at which we see divergences at~all.
In the same way, the theory for $\gamma=1$ we call 3-loop super-renormalizable
in $d=4$ dimensions. If we have UV-divergences only at the one-loop
level, then we are in a much better situation in hoping for cancellation
of them in a some model of UV-finite theory.

There is already a known approach to QFT, which gives the desired result
of QFT without any beta functions. Actually, the condition of vanishing
of beta functions is equivalent in the parlance of exact RG (or functional
RG termed also as FRG) community to the existence of a fixed point  of Renormalization
Group flow (RG), which is true at all energy scales.
This means that the Wilsonian effective action of a theory should
always stay at the fixed point, not only in the UV regime. We look
for UV-finite theories such that in the UV regime they meet a non-trivial
fixed point (meeting a trivial fixed point is quite of boring, because
this is a free theory without any gravitational interactions). This
is one of the conditions present in the asymptotic safety programme
(AS) for quantum gravitational interactions \cite{as1,as2,as3,as4}.

One of the first papers with explicit construction of UV-finite quantum gravitation
with higher derivatives appeared in 1989 by Kuzmin \cite{kuzmin}. The idea by Kuzmin was the
pioneer one and is still very interesting. It has moreover some relations
to models with fractional derivatives \cite{fractional1,fractional2,fractional3}.

As a first element in the last step of construction of UV-finite quantum
gravitational theory, we remind how the general structure of perturbative
UV-divergences looks at the one-loop level. The divergent part of
the effective action at this level can be written in a form:
\begin{equation}
\Gamma_{{\rm div}}=\!\int\!d^{4}x\sqrt{|g|}\left(\beta_{R^{2}}R^{2}+\beta_{C^{2}}C^{2}+\beta_{E}E\right)+\left(\beta_{\square R}\square R\right)\,,\label{eq: divpart}
\end{equation}
where $E$ above is another notation for the Gauss--Bonnet scalar (since
it gives rise to Euler invariant of the four-dimensional manifold)
and we have included also an additional term $\square R$, which is
clearly a total derivative. One may ask why we need to include that
Gauss--Bonnet term and the last one to the divergent effective action
at the one-loop level. Actually, it is perfectly true that they do
contribute nothing to the EOM of the effective theory (to their divergent
part at one loop); however, the inclusion of such two terms, which after variations
are total derivatives, is useful, when we want to discuss the general
symmetry properties of the theory. This is important because, when we want to
perform some symmetry transformations to check the invariance of the
action, then these terms may be essential in providing the correct
answer as it is, for example, in some cases in supersymmetric theories.
However, here we will be mainly interested in conformal symmetries of the
theory and for them the expression above gives rise also to the correct
and complete expression for the conformal anomaly of the theory, which~we will discuss
here in short. All terms in the action above contributes to the conformal
anomaly and forgetting about one of them would spoil the derivation
related to conformal symmetry at all. 

To kill these divergences, we will use killers. What are killers? In
the construction of a higher derivative theory (which could be understood
as a viable UV limit of some non-local theory with form-factors) so
far we have considered only operators, which were quadratic in field
strengths (gravitational curvatures). However, it is possible to add
also other operators, if they do not change the renormalizability and
super-renormalizability properties
of the theory. For this to be true, we need the number of derivatives
on the metric in additional operators to be smaller or equal
to the number of derivatives, which are in terms quadratic in curvatures,
which completely determine the form of the propagator on flat spacetime.
Then, the renormalizability (and super-renormalizability too) is safe
and moreover we know, which operators in the action contribute to the
flat spacetime propagator. (Of course, in the full non-local theory,
we can add non-local killer terms to the action that only in the UV
regime attain the form of higher-derivative terms, which respect the
renormalizability, and only in this limit have the number of derivatives
bounded by the one appearing in the inverse of the graviton propagator
in the UV limit.) The killer operators have an impact on the beta functions
of the theory, but~not on the graviton propagator around flat spacetime.
They are among the other group of terms, which still could be added to
the action and this change does not modify super-renormalizability
properties of the theory. One can also speak about terminators and
spectators. The terminators participate in killing the beta functions
of the theory, but they also modify the propagator; however, the way they
are written hides their true nature because they do not look like
terms obviously quadratic in gravitational curvatures. The last criminals,
which appear at the crime place (of killing the beta functions), are spectators.
These operators, still allowed by super-renormalizability, do not contribute
neither to the propagator, nor to the beta functions of the operators
considered above. They will surely contribute to the finite terms
of the effective action at one-loop, but for study of divergences
we do not see any effect of them.

Focusing on killers, now we can easily understand that they can be
at most quartic in curvatures in four dimensions. This is a corollary
of the fact, how the expression of the effective action at one loop
looks in an abstract way
\begin{equation}
\Gamma^{(1)}=\frac{i}{2}{\rm Tr}\ln\frac{\delta^{2}S}{\delta\phi^{2}}\,,\label{eq: oneloopeff}
\end{equation}
where an important fact is that the functional trace denoted above
by ${\rm Tr}$ is a linear operation and we will expand the natural
logarithm according to the power series:
\begin{equation}
\ln(1+x)=\sum_{i=0}^{+\infty}(-1)^{i+1}\frac{x^{i}}{i}\,.\label{eq: logexp}
\end{equation}

 Since we
know that the ansatz for the divergent part of the effective action
in four dimensions has the structure of terms, which are at most quadratic
in gravitational curvatures Eqn. \eqref{eq: divpart}, then we easily understand by the argument
presented above that the terms which are quintic and higher in curvatures
will not contribute anything to this divergent action $\Gamma_{{\rm div}}^{(1)}$.
The killers can be at most quartic in curvatures. (This result depends
on the dimensionality of spacetime and in general dimension $d$ the
killers for one-loop beta functions may contain up to $d+2$ covariant
curvature tensors.) Therefore, they can be cubic or quartic in curvatures,
while at the same time they may contain any even number of covariant derivatives acting
on these curvatures (but these terms as whole must respect the renormalizability criterion
from the above).

Due to the mentioned linearity, let us choose to use quartic killers
only and use the following addition to the original action:
\begin{equation}
S_{{\rm kill}}=s_{1}R^{2}\square^{\gamma-2}R^{2}+s_{2}C^{2}\square^{\gamma-2}R^{2}+s_{3}E\square^{\gamma-2}R^{2}+\label{eq: killers}
\end{equation}
\[
+\left(s_{4}(\square R)\square^{\gamma-2}R^{2}\right)\,.
\]

It is easy to understand how our three (four) killers are constructed.
We took each term in the ansatz for the divergent part of the effective
action at the one-loop level in Eqn. (\ref{eq: divpart}) and simply multiplied them by the scalar
expression $\square^{\gamma-2}R^{2}$, where the box operator is understood
to act on the square of the Ricci scalar on the right. We also multiply
each term by a suitable coefficient $s_{i}$, whose energy dimension
is the same of the coefficients $\alpha$'s with the highest $N$ appearing in the quadratic
part of the higher derivative action Eqn. (\ref{eq: hihider}). This also
ensures that the killers in Eqn. (\ref{eq: killers}) are not in conflict
with renormalizability of the whole theory. There is a mentioned advantage
of using quartic over cubic killers in four spacetime dimensions because
in the final contributions for beta functions the coefficients $s_{i}$
appear only linearly.

Indeed, in bigger generality, the expressions for all beta functions
of the theory we write in the following form:
\begin{equation}
\beta_{i}=v_{i}+a_{ij}s_{j}\,.\label{eq: systemeqns}
\end{equation}

This displays a linear structure of the system of the beta functions of the theory.
The matrix $a_{ij}$ is some real constant matrix. The contributions
collectively put in $v_{i}$ denote all contributions to the beta
functions from all other terms different than four selected killers in the action (from possibly other killers,
possibly also cubic and also from terms quadratic in curvature in the action). These are
all contributions independent on the coefficients $s_{i}$ of our
four chosen killer terms Eqn. \eqref{eq: killers}. The second term in the above sum in Eqn. (\ref{eq: systemeqns})
is obviously linear in the coefficients $s_{i}$. Now, the question
arises, but~it~is simple to answer: how we can make our theory so
special that it will have no beta functions at all? This~is equivalent
to demanding from the theory to be so fine-tuned that it is conformal
on the quantum level. The condition for UV-finiteness is simply
\begin{equation}
\beta_{i}=0\,.\label{eq: fincond}
\end{equation}

The answer is a matter of simple algebra, since we can always find
solutions for unknown $s_{i}$ coefficients, if the $v_{i}$ contributions
are all given and if the matrix $a_{ij}$ is non-degenerate. This
is after all a system of linear equations and, provided that the matrix
$a_{ij}$ is non-degenerate, we can always find a real solution
for the unknowns. The non-degeneracy of the matrix $a_{ij}$ can be
easily grasped as follows. This is the 
result of the detailed computation of the second variation of these terms
 with respect to gravitational fluctuations. 
As from the structure of what is left after stripping $R^2$ from these terms,
 one can see that there is a contribution to each of the terms
 present in the divergent part of the effective action Eqn. \eqref{eq: divpart}. 

Now, the solutions to the linear system of
equations Eqn. (\ref{eq: systemeqns}) always exist and we denote it collectively
as coefficients $s_{i}^{*}$. According to what was said above
the HD theory:

\[
S_{{\rm UV-fin}}=\!\int\!d^{4}x\sqrt{|g|}\left\{ \vphantom{\left(s_{4}^{*}(\square R)\square^{\gamma-2}R^{2}\right)}\alpha_{R}R\square^{\gamma}R+\alpha_{C}C\square^{\gamma}C+\right.
\]
\begin{equation}
\left.+s_{1}^{*}R^{2}\square^{\gamma-2}R^{2}+s_{2}^{*}C^{2}\square^{\gamma-2}R^{2}+s_{3}^{*}E\square^{\gamma-2}R^{2}+\left(s_{4}^{*}(\square R)\square^{\gamma-2}R^{2}\right)\right\} \label{eq: UVfinact}
\end{equation}
is completely UV-finite in four dimensions for $\gamma\geqslant3$
and for precisely tuned values of the coefficients $s_{i}^{*}$, which
are the results of solving the system of equations from above Eqn. (\ref{eq: systemeqns}).

In the example above, we have found a candidate theory for fundamental
Quantum Gravity, which is a UV-finite theory. Hence, it is also conformal
on the quantum level. Since there are no divergences beyond the one-loop
level, in the theory, there are no divergences at all and conformality
is completely preserved by quantum corrections. The achievement above
is equivalent to finding a theory, in which conformal anomaly to all
loop orders is completely absent. We showed explicitly that, at the
one-loop level, there are no UV-divergences and the theory is with
all beta functions set to zero. In QFT, the trace of the energy-momentum
(pseudo-)tensor of the system derived from the effective action
is proportional to the beta functions
of the theory, or in other words to the conformal anomaly of the theory.
We have explicitly in formulas:
\begin{equation}
T=g^{\mu\nu}T_{\mu\nu}=\sum_{i}\beta_{i}{\cal O}_{i}\,,\label{eq: trace}
\end{equation}
where the full action of a system we write as
\begin{equation}
S=\!\int\!d^{d}x\sqrt{|g|}\sum_{i}\alpha_{i}{\cal O}_{i}\label{eq: genact}
\end{equation}
and the beta functions $\beta_{i}$ are the logarithmic derivatives
with respect to energy scale $k$ of the running coupling coefficients
$\alpha_{i}=\alpha_{i}(k)$ (some of them run, not necessarily all).
The set of operators ${\cal O}_{i}$ consists of all generally and
gauge covariant operators that we want to add in the Lagrangian of
the theory. They~have to respect all symmetries assumed from the theory
on the classical level. The~non-vanishing of the trace on the quantum
level of the energy-momentum tensor is called a quantum trace or conformal
anomaly. It arises in theories which are conformally-invariant on
the classical level and for which the process of quantization produces
scale-dependence of couplings and in the result quantum trace is non-zero.
It is well known that, for normal theories, conformal invariance on
the classical level is just an accident and it is generically removed
by quantum corrections, which force couplings to run (like for example
in YM theory). The theory must be very special to avoid conformal
anomaly. However, since in the standard framework of QFT we know how
to live with theories that predict and indeed exhibit RG running,
then typically this non-vanishing anomaly is not problematic.

For gravity, the situation is diametrically different, since we want to gauge
this symmetry (make it in a local version).
Therefore, making conformal 
symmetry non-anomalous is very crucial in
construction of a good quantum gravity model. In our case, the physical
system is a quantum self-interacting gravitational field. Hence, we
cannot truly speak about the energy and momentum of gravitational
field here, since they are not generally covariant notions in E--H general
relativity. However, some covariant parts of the energy-momentum pseudo-tensor of relativistic
gravitational field could be still defined and transform well
in some higher derivative gravitational theories~\cite{highder}.
 This is why
in the pure gravitational setup we should speak only about the trace of the energy-momentum
pseudo-tensor.

The trace of the energy-momentum pseudo-tensor of the gravitational field at one-loop
level in the gravitational theory in four dimensions off-shell \cite{confan,confan2,confan3,shapiroconf1,shapiroconf2}
takes the following simple form:
\begin{equation}
T=\tilde\beta_{C^{2}}C^{2}+\tilde\beta_{E}E+\tilde\beta_{\square R}\square R\,,\label{eq: tracean}
\end{equation}
where all three coefficients $\tilde\beta_i$ are linear combinations of the original beta functions $\beta_i$
of the theory as expressed in the divergent part of the effective action in Eqn. (\ref{eq: divpart}).

Here, we eventually see why it is important to keep the last two terms
in the divergent part of the effective action Eqn. (\ref{eq: divpart}).
The expression for the trace anomaly is not under a spacetime integral;
hence, we~cannot integrate by parts and we cannot neglect terms with
total derivatives here. The last two terms are generally non-vanishing
on curved background manifolds
and should be included for the proper account of the conformal anomaly.
However, in our finite theory, we took care of them too (by~adding
two suitable killers for $\beta_{E}$ and $\beta_{\square R}$). By killing all four
beta functions in Eqn. (\ref{eq: divpart}), we also set to zero any linear combination
of these coefficients, so, in particular, the combinations appearing in Eqn. \eqref{eq: tracean}.
In the result, the total trace of the gravitational
 pseudo-tensor vanishes in finite theories. This is another check of conformal consistency
achieved on the quantum level.

Finally, it is necessary here to mention the following situation
happening in UV-finite theories based on super-renormalizable
gravitational theories. Such theories, as exemplified by the action Eqn. 
\eqref{eq: hihider}, contain dimensionful parameters in their higher
derivative parts (we neglect here the issue with the presence of the Newton's
and the cosmological constant, although they are obviously dimensionful too).
This signifies that on the classical level such theory cannot be
scale-invariant and exhibits a classical contribution to the trace
anomaly $T_{\rm cl}$ of the gravitational energy-momentum pseudo-tensor.
This contribution $T_{\rm cl}$ is related to the presence of all these
dimensionful couplings, and not to their RG running, since there is
no RG running on the classical level of the theory. Now, inclusion
of quantum corrections adds a new quantum contribution to the anomaly
$T_q$. What we denote in the formula Eqn. \eqref{eq: tracean} is a total
contribution to the conformal anomaly $T=T_{\rm cl}+T_q$. We see that in UV-finite
gravitational theories, the total trace vanishes, or in other words
the contributions from quantum loops (which are true RG effects) $T_q$
cancel classical contributions $T_{\rm cl}$ spoiling
scale-invariance on the classical level of the theory. This interplay
between classical and quantum effects is a characteristic feature
of HD UV-finite gravitational theories.
Moreover, here we have 
a cancellation of tree-level classical features
not only by  one-loop quantum effects but by all loop quantum effects, since
the theory for $\gamma\geqslant3$ is one-loop super-renormalizable.
That is, this effect and the virtue of UV-finiteness cannot be reverted
 by a situation at the two-loop level.

Concluding, we are able to write local conformal
symmetry as a consistent symmetry of the gravitational interactions
on the quantum level. The fact that all the beta functions vanish
we can see as a reason for scale-invariance of all Green functions
(for any number of legs, on-shell and off-shell, as~well as for any
number of loops). This determines that the dynamics of the theory
is indeed governed by the conformal symmetry on the quantum level.
Just obtained Conformal Quantum Gravity stays always at the non-trivial
fixed point of the RG. In addition, on the other hand, in this Quantum Gravity, there
are completely no divergences, so the problem with infinities is finally
solved and the theory fulfills old expectations from the old time,
when QFT was born. The long story of infinities is finished.

\section{Conformal Quantum Gravity\label{sec:CQG}}

Having constructed a first viable candidate theory for higher derivative
UV-finite quantum gravity, we can now come back and analyze in short
the question of why we need a conformal field theory (CFT) of quantum
gravity. Besides the obvious enhancement of symmetries of the theory
and their preservation on the quantum level, we notice that now this
model of UV-finite theory is amenable for answering any problem previously
associated with quantum gravity. This theory possesses a very good behaviour
at the high energy scales (UV limit) and, due to the non-local structure of the theory,
this good behaviour is extended to any other scale, and for example
the theory is ghost-free. As~such, this CFT is a basis for doing conformal
perturbation theory, when some operator(s), which is not conformally invariant,
is added to the theory and causes the RG running. We know that, in
the real world, conformal symmetry must be broken and this is one way that
this breaking can be achieved: by adding some non-trivial operator
constructed from CFT data as a perturbation to the fixed point behaviour. The
symmetric phase of the theory with unbroken conformal symmetry is
the basis for studying of various deformations. 

The problem of coupling matter to CQG is also interesting. We would like to shortly remark
here that quantum aspects of conformal higher derivative quantum gravity
 coupled to conformally invariant matter were studied in \cite{Eli1,Eli2}.
In addition to this, the generalization of conformal higher derivative gravitational theories 
for the presence of scalar was done in \cite{Eli3}.
Finally, the $4-\epsilon$ expansion RG technique was applied to conformal $C^2$ gravity
with conformal matter in \cite{Eli4}.

Moreover, in conformal
field theory, we have also (trivial) solutions to the unitarity problem
because, in such theory, the scattering matrices are not defined
and there is no any problem of unitarity violation. This is, when we
insist that the situation with asymptotic states must be conformally
invariant too. We remind readers here that, by unitarity of the physical theory,
we really mean the unitarity of the $S$-matrix in such theory. Hence,
in theories, where we are unable to properly formulate the problem
 of scattering, we cannot speak about the unitarity of scattering.
 This is, for example, a situation met on non-trivial spacetime
 backgrounds (like in cosmology),
 which do not possess asymptotically flat regions.
Basically, in the unbroken (symmetric) phase of the conformal symmetry,
 we similarly cannot properly define asymptotic states. Every trial to define them leads
to inevitable breaking of conformal symmetry and this is a breaking
by exact solution of the classical theory. If there are no problems
with unitarity and the scattering matrix is trivial, then this also
signifies that there is no putative breaking of the unitarity bound
for the scattering amplitudes. This is, of course, related to the preservation
of the full unitarity in the theory. We also discuss the issues with
scattering amplitudes in CFT below in the next section.

On the other hand, the condition on the theory to be conformal puts
very strong constraints on the theory and the effective action of
it. We can say that, for example in $d=4$, this determines the conformal
 theory uniquely, if there are no mass scales in the theory. In other cases, the constraints enforced
by conformal symmetry are so strong that they restrict the possible
form of the form-factor in non-local theories. The constraints are
not only put on the terms in the effective action, but also on the
anomalous quantum dimensions of the operators. Since Green functions
of the theory do not show any scale-dependence, then the only non-trivial
elements of QFT are these anomalous dimensions of the operators here, since couplings
in front of operators cannot run. There is no RG running, which could
be read from the divergences of the effective action. On the quantum
level of the theory, we only have finite renormalization of couplings
and no scale dependence. The bare values of the couplings that we
put in the original classical Lagrangian of the theory generically
are not the same as constant values, which appear in the effective
action. This is the only difference between classical and quantum
effective action for conformal theories, which are without any massive
parameter on the classical level. However, this finite renormalization
of couplings has nothing to do with UV-divergences of the theory and
it is just a requirement that our theory takes the values of the couplings
from the experiment to make a relation to reality. In principle, the bare
action with bare couplings has no relation a priori to couplings that we meet in Nature.

Moreover, since there are no divergences in the theory, we do not need
to put and add any renormalization scale (typically denoted by $\mu$),
which is an optional parameter with one mass dimension, which is typically
introduced in non-conformal theories on the quantum level for the purpose
of renormalization of infinities in the effective action.
This is very great news, since the number and the character of the
classical parameters of the theory is preserved and there is no addition
of this scale $\mu$. Actually, one can look back and easily understand
that the big diversity and complicacy of the effective actions that
are found at any loop level in non-conformal theories is due to the
appearance of this scale $\mu$ and the possibility to construct infinitely
many various terms with it in the effective action. When $\mu$ is
not needed, then we do not use it for the construction of a quantum
effective theory and the effective action must look much simpler.
Finally, as mentioned in the introduction, the presence of the conformal
symmetry on the quantum level signifies that the quantization procedure
is idempotent and that the quantum effective action is uniquely determined
and takes the same structural form as the original classical action
of a theory.

Let us also make some detour here and analyze and describe a different
model (with conformal dilaton field), which explicitly shows that
the conformal symmetry is present both on the quantum as well as on
the classical level of the action of UV-finite quantum theory. This
way, we can make any UV-finite theory manifestly conformally invariant
even on the classical level \cite{dilaton}. We need to add a spurious (that is a
non-dynamical) additional field to theory, which we can technically call
a conformal compensator field $\phi$. The more common name for this
field is dilaton, although, for example, in string theory, it plays a
slightly different role and there it is a dynamical field. We can
achieve this by performing the following substitution in the gravitational
part of the effective action, which we treat as a functional of the
metric tensor $g_{\mu\nu}$ in general dimension $d$:
\begin{equation}
g^{\mu\nu}=\left(\phi^{2}\kappa_{d}^{2}\right)^{\frac{2}{2-d}}\hat{g}^{\mu\nu}\,.\label{eq: confsubst}
\end{equation}

This is reminiscent of how we can perform conformal transformations
on the metric tensor \hspace{1cm} ($g_{\mu\nu}\to\Omega^{2}(x)g_{\mu\nu}$)
 and what is the energy dimension of the parameter $\kappa_d$. We
want the field $\phi$ to have the single energy dimension in spacetime
of any dimensionality $d$. Moreover, in the formula above, we denote
by $\kappa_{d}^{2}$ the inverse of the square of the $d$-dimensional
gravitational Newton's constant. The~dilaton compensates over the
conformal transformations of the metric, if itself it transforms under
conformal transformations according to the following rule:
\begin{equation}
\phi\to\Omega(x)^{\frac{2-D}{2}}\phi\,,\label{eq: diltr}
\end{equation}
where here $\Omega^{2}(x)$ is a general parameter of the conformal
transformations. The factor $\Omega^2$ appears also when the conformal transformation
acts on the metric tensor $g_{\mu\nu}$ with both indices
covariant. An action for a UV-finite theory with the metric $g^{\mu\nu}$
substituted by $\hat{g}^{\mu\nu}$ as in Eqn. (\ref{eq: confsubst}) is
not only conformal on the quantum level because all of the beta functions
vanish, but also is manifestly conformally-invariant, if written entirely
in terms of the metric $\hat{g}^{\mu\nu}$. This is because the metric tensor
$\hat{g}^{\mu\nu}$ is invariant under conformal transformations.
Of course, the first condition here is that
all perturbative beta functions of the theory vanish at any energy scale. Only
under this condition can we perform and only then the dilaton trick works.

In such theory, we have
algebraically ''9 + 1'' degrees of freedom in the metric field $\hat{g}^{\mu\nu}$,
since the last tenth degree of freedom is taken by the dilaton field
$\phi$. This is regarding the algebraic counting of degrees of freedom,
while the dynamics of the theory is, of course, very different and the total
number of degrees of freedom and their character (and spin) is typically
lower and maybe different from these algebraic considerations. (To give
a basic example: in E--H theory algebraically, we also have 10 degrees
of freedom in a symmetric metric field in $d=4$; however, there are
less physical degrees of freedom in the dynamics, and they could be
identified with only two helicities of the massless graviton particle.)

Moreover, as emphasized extensively in other papers on the series of
conformal symmetry, this symmetry is crucial in solving the problem
of classical singularities of classical gravitational theories. A
proper choice of the conformal scale factor $\Omega^{2}(x)$ makes
the GR-like singularities gauge-dependent on the conformal gauge.
In addition, as we know, in the true physical description of the system, we cannot
use information (like invariants, etc.), which do depend on the gauge
choice, and which would kill the gauge symmetry transformation of the
theory. Since about some singularities of classical gravity theories
we conclude by analyzing the classical non-conformal invariants, then
these statements have no meaning in truly conformal gravity. In addition, this
regards the singularities appearing in the exact solutions of the
classical theory, as well as in solutions of the effective theory arising
on the quantum level. Conformal symmetry is crucial in making the
theory free of any classical pathologies, not~only singularities,
but, for example, of closed time-like curves (CTC's), which do appear
in such vacuum solutions of Einstein equations as for example found
by G\"{o}{}del. This pathology is resolved by conformal scale factors,
which make the time lapse of such curve not a conformal gauge-independent
observable and hence it effectively means that any causality breaking
effect related to such loop does not appear. Effectively, we can say
that the conformal factor rescales the time lapse to zero on such
curve and being on it we cannot run into any problem. Similarly, use
of conformal transformation resolves various topological singular
objects and regions of spacetime, like for a singular ring of Kerr classical
spacetime. As it was emphasized earlier, the issue of UV-divergences
of a theory is intimately, but non-intuitively, related to the issue
of classical small-distance singularities of the theory. This~correspondence
is also very non-trivial because it links perturbative phenomena of
quantum theory with highly non-perturbative strong coupling regime
phenomena of the classical theory. In its rough structure, this scheme
resembles by now very famous gauge/gravity duality.

We also would like to remind readers that, in the real world, the conformal
symmetry must be present in the broken phase and this we can achieve
in different ways. One of them was previously introduced with adding
some non-conformal operator to the action of CFT and therefore causing
a non-trivial RG flow. Another possibility is to exploit, so popular
in particle physics, spontaneous breaking (SSB) of conformal symmetry.
Since here the conformal symmetry is in the local (gauged phase),
then the proper mechanism to achieve this is a conformal gravitational
analogue of the Higgs mechanism. In~this way, we assume that vacuum
expectation value of the dilaton field in the state of the true vacuum
of the theory is non-zero and this {secretly} breaks conformal symmetry.
Although it is known that the symmetry, which is local is never truly broken,
since it is exact and not approximate, just the Fock state of the vacuum
of the theory hides some of all possible symmetries of the theory.
 We achieve this, if we require that in the physical
phase we have: $\langle\phi^{2}\rangle\neq0$. This condition for
the dilaton field introduces a new mass scale to the theory, since
the field $\phi$ has the single power of the energy dimension. Such~scale was not present in the original action of the theory. It is
in all computations regarding the theory in the broken phase because
it comes with the choice of the vacuum state. These issues of the
breaking of the conformal symmetry are strictly related to the question
about the vacuum state in conformal gravitational theories.

Finally, we would like to admit that conformal gravity (both on the
quantum as well as the classical level) is a theory, which fully realizes
Mach ideas about the spacetime and matter and their interrelations.
We remind that Einstein was influenced by Mach; however, in his standard
GR theory, he did not include Mach's profound ideas completely. In Einstein's
theory, the inertia of particle (and~in bigger generality its energy
content) does not depend on the constitution and configuration of
spacetime out there, away from the point of the actual location of
the particle. Conformal gravity is on the contrary very Machian and
this idea is fully embodied by the study of conformal transformations
of the metric. The possibility of freely doing them, since they are
symmetry transformations of the theory, is the link which relates actual
location of the particle (here) with global structure of the spacetime
(out there). Moreover, the ideas of conformal and Machian origin of
mass are very influential for the understanding of the Higgs mechanism
and the gravitational origin of mass in the SM of particle physics.
It is another virtue that conformal symmetry may help in unifying
a well defined UV-finite theory of quantum gravitational field with
a particle physics and in providing the same theoretical grounds for origin
of both gravitational mass (related to gravitational interactions)
and inertial mass (related to other interactions in particle physics,
and to Higgs phenomena especially). We are hoping that (probably) a full unification
of all fundamental interactions due to conformal symmetry is within
our reach.

\section{Conformal Symmetry of the (One-Loop) Effective Action\label{sec:Conformal-symmetry-of}}

As we know in quantum field theory, if there are no divergences of
the perturbation theory, then there are no perturbative beta functions
of running couplings. We want to discuss the situation at the one-loop
level only, since at higher loops the super-renormalizability properties
of some theories are only conjectured and they are based on power-counting analysis
of UV-divergences. Such overall (or~naive) analysis may be hampered
by the fact that two or more loop integrals have to be done, even~in the
Euclidean signature. In addition, as it is known from the studies for the Minkowskian signature
case, new~UV-divergences may show up after doing the second integral
and the analysis of powers for these divergences may not give the
expected results. However, there is a rescue for super-renormalizability
of higher derivatives theories at higher loop orders in Euclidean
signature and the general proof (not~based on power-counting analysis)
can be done using, for example, Batalin--Vilkovisky formalism of quantization.
We also remark that the whole notion of super-renormalizability is
tightly related to power-counting and, beyond this, it does not make
any sense. Therefore, we will focus on the situation at the one-loop level,
treating this first term in the expansion in powers of coupling constant
as a physically meaningful one. This is also an expansion in $\hbar$---the quantum physics constant. Of course, since we will not have
typically an access to a full resummation of all perturbative contributions
or even to non-perturbative contributions, such terms in the expansion
still provide us with a very useful information about the quantum
physics of some quantum field theory models.

In UV-finite theories, we have a special situation. The question is
how special. The~analysis here will be independent from the particular
regularization scheme used to handle with UV-divergences. We can obtain
beta functions using dimensional regularization, cut-off regularization
or even $\zeta$-regularization on curved spacetimes. Maybe this analysis
here will not apply only to exact beta functions obtainable, for example,
in Exact Renormalization Group approaches to QFT (because we do not
know how to make all such beta functions vanish by our standard tools).
Here, we will consider other examples of theories, which are UV-finite
at the one-loop quantum level. However, as described in Section \ref{sec:Story-of-infinities},
the UV-finite gravitational theories that we found there are finite
with all loop orders accuracy because they are one-loop super-renormalizable.
However, for the sake of simplicity, here we will also study theories of different
types, which firmly shows finiteness at the one-loop level. We know that, in these
models, there are no divergences and beta functions are zero. What
does this imply for one-loop effective action? The answer for the
divergent part of this object is very simple. This is zero in finite
theories. Again, we emphasize that there is no divergent part of the
effective action at one loop. This we can achieve for sure and this
is what we have seen so far.

Since beta functions are zero, we can conclude that the conformal symmetry
is somewhere there at the one-loop level. We may say very modestly that
it is in the divergent part of the effective action, but~this is a trivial
fact, since this part of the action is identically zero. Thus, of course,
it is there and infinitely many of other symmetries are also enjoyed
by the divergent part because zero homogeneously always transforms
back to zero. What does it mean and what is the significance of this
for the full quantum theory at the one loop? Does this mean that the effective
action (finite terms) possesses the conformal symmetry or any other
symmetry? Such conclusion is obviously too fast and seems to be an
absurd one. We know that there is also a finite part of the effective
action and this part often will not enjoy symmetries in question.

Let us be a bit more clear about this point. There is a lore of QFT that
all the symmetries of the divergent action must be also enjoyed by
the finite terms of the effective action. However, since there is no a
rigorous proof of such assertion, we are here in position to question
it. Does it happen always? What if the divergent action is zero, like
above? We cannot derive from this that effective action enjoys any
symmetry, which may come to our mind. This is not true, as far as we
know, for some effective actions in some QFT models. We think that the
significance behind this not rigorously stated lore is that, in good
QFT models, quantum anomalies should not arise exclusively due to finite
parts only. If there is anomaly of some symmetry, which is by the
way enjoyed by the classical original action, then it should show
up already in the divergent part and cannot be only visible in the
finite part. However,~this regards the anomaly of the symmetry of the classical
action and not a symmetry of the quantum action! Typically, the quantum
theory cannot enhance its symmetries beyond the level which was seen
classically. Now,~with UV-finite theories, the symmetry seems to be
enhanced at the quantum level. For sure, there is some enhancement and
some new properties of such UV-finite theories, but, for the moment,
it is too much to say that there is a conformal symmetry there. We
will argue about this also below. However, at least from this paragraph,
it is clear that we cannot use the wisdom of QFT blindly, saying that
any symmetry enjoyed by the divergent action must be also respected
in finite terms.

Having no divergences is quite a special property of the effective action.
However, conformal symmetry of the full effective action at one loop seems
to be even more special. Here are the arguments. Let us consider first
the situation in those higher derivative theories of gravitation,
where there are many tensorial terms that we can construct and add
to the gravitational action not changing at all very good UV properties
of theories.

Let us consider first the term like ${\cal R}^{5}$ in four spacetime
dimensions added to a first one-loop super-renormalizable theory (with $\gamma=3$). The
action of a main theory we schematically write as
\begin{equation}
S_{{\rm grav},4D}=\!\int\!d^{4}x\sqrt{|g|}a{\cal R}\square^{3}{\cal R}\,.\label{eq: 4daction}
\end{equation}

It is obvious (due to Barvinsky--Vilkovisky formalism \cite{BVreport}) that the term ${\cal R}^{5}$
does not contribute to divergences of the theory at the one-loop level.
This happens, provided that the number of derivatives on metric in
such term is not higher than the number of derivatives on metric in
the term in Eqn. \eqref{eq: 4daction} that shapes the UV behaviour of the gravitational propagator,
so the renormalizability of the theory is not spoiled. If the original
theory (without this  ${\cal R}^{5}$ term) was finite, then adding this term with
whatever front coefficient to the Lagrangian does not change this
fact about divergences. In addition, for terms containing also covariant derivatives or
 higher terms in curvature, this the same (in a situation,
 when also the exponent $\gamma$ is bigger than the minimal $\gamma=3$). Simply,
such terms are spectators from the point of view of killing the beta functions. Do
we really think that the putative conformal symmetry of the effective
action could be so robustly independent on values of so many (in principle
very many, only constrained by renormalizability requirements) couplings
in front of terms like ${\cal R}^{5}$ and higher in curvatures? Then,
this would not be a very special symmetry but a robust feature! Such
is the UV-finiteness though.

We think that conformal symmetry should be more constraining, like
it is on the classical level. There are only few examples of classical
actions with conformal symmetry and they are constructed only with
a really small number of terms of lower order in curvatures. Coming
back to the impact of such after-killer operator, it cannot have any
influence on the finite terms of the effective action of the form
${\cal R}{\cal F}_{2}(\square){\cal R}$ either. (This is because of the trace of
the second variation, which gives the effective action at one loop.)
However, it can influence the terms of the type ${\cal R}^{3}{\cal F}_{3}(\square){\cal R}$
and higher in the effective action. These last terms are not in any way linked to
the UV-finiteness, but they could be important for a construction of
an effective action enjoying full conformal symmetry.

Actually, following
some ideas suggested in FRG
framework for RG flows in QFT, we can
think of the finite pieces of the effective action as of a series that we have started to build for a conformal
theory. The divergences are not there, but this fact is linked to the
specific forms of terms of the types ${\cal R}{\cal F}_{2}(\square){\cal R}=O({\cal R}^{2})$,
$O({\cal R}^{3})$ and $O({\cal R}^{4})$. However, then, when we include
the ${\cal R}^{5}$ term to the original action, some terms in the
effective action of order ${\cal R}^{3}$ and higher have to be modified
accordingly because there is an impact of such operator on these parts of the effective
action. In addition, now, if originally the effective action (finite pieces)
was conformal, then addition of such operator (of the type  ${\cal R}^{5}$) may change something in this series,
in order to recover the conformality we may need to readjust (or even to determine) the coefficients
in front of ${\cal R}^{5}$ terms. In this way, we construct step by
step a series giving the finite parts of the effective action, possibly
enjoying conformal symmetry, when new higher in curvature operators
are added. We cannot exclude now the possibility that this will lead
to conformally invariant effective action. However, we are sure that,
for the divergent parts in the UV regime (with beta functions), the
series truncates after few initial terms and we do not need to go
to the next step in reconstructing full UV-finite theory. If the theory
is a minimal UV-finite one (without operators of the type ${\cal R}^{5}$),
then for the finite pieces of the effective action at one-loop, we
nonetheless have a series of operators $O({\cal R}^{2})$, $O({\cal R}^{3})$,
$\ldots$---the so called curvature power expansion of the effective
action. The~coefficients in this series have to be computed sequentially,
so again we can think that the series for conformal theory is possible
to be constructed, but at the same time it is very uniquely determined.

Now, still in the super-renormalizable theory Eqn. \eqref{eq: 4daction} at one loop, let us concentrate
this time on the impact of the operator ${\cal R}^{2}$ (here precisely without
 covariant derivatives), when added
to the original action Eqn. \eqref{eq: 4daction}. Again, this does not contribute anything to
the UV-divergences. However, this time, because it has too few derivatives on the metric, but not because of too big number of
curvatures (like it was in the previous case), it does not influence
completely UV-finiteness and perturbative beta functions. However, still
it influences the effective action (finite terms) at one loop. We
can even say to which term it contributes to the term ${\cal R}{\cal F}_{2}(\square){\cal R}$
and possibly higher in curvature too. Again, it can come with whatever
coefficient, but finiteness is very robust and independent of it. Do
we think that the supposed conformal symmetry will also be so insensitive
and will allow for such a big parameter space? Actually, here there
is a finite number of operators with such small number of derivatives
on metric and small powers of curvatures (for example, the E--H term is in such a set).
 However, if the conformal symmetry
would let this happen, then we do have very little constraints
on the form of the conformal action. We think it is not the case and
conformal symmetry cannot be so robust. Moreover, the term ${\cal R}^{2}$
with any coefficient contributing to the effective action will spoil
the series of supposed conformal action, that we described and constructed
above. (We remind that this series starts with a term of the type ${\cal R}\square^{3}{\cal R}$ in the original classical action of a theory.)
If there is a conformal symmetry there, it should constrain
the impact of the term ${\cal R}^2$ somehow, but we do not see any constraint
like this on the level of the perturbative beta functions.

Finally, we want to bring the last argument, why the UV-finiteness
is so robust, but conformality should not be. This is when we view
higher derivative theories only as UV limits and asymptotics of truly
non-local theories with some form-factors. For UV-finiteness, only
the behaviour in the UV regime matters, so we concentrate on higher
derivatives only, like we did in the two previous paragraphs. This
UV-finiteness is again completely independent from the structure of
form-factors at finite energies. If we achieve UV-finiteness for one
asymptotics of form-factors, then this is a robust feature enjoyed by
all form-factors with the same UV behaviour. However, still form-factors
at the middle energies may be very different and the effective actions
(finite pieces) must depend on these differences. However, we notice that
there exist some gravitational observables, which completely do not
depend on the precise analytic form of form-factors. These are, for example,
tree-level gravitational scattering amplitudes~\cite{scattering}. Conformal symmetry
should be very special, but it does not constrain form-factors and
in this case the possible parameter space is extremely huge because
of the choice of possible interpolating form-factors as analytic functions.
The UV-finiteness is a very robust and easy to achieve property
of some higher derivative or non-local models, but conformality should
be a very special property, so it still should be very different from
UV-finiteness, but should require the former as the first necessary
step. In other words, it is very improbable and unlikely that the theory,
which is UV-finite, so from quite a huge class, will reveal itself to be a conformal
one, which must be a very restricted and a special theory. For general
UV-finite theories, this is like a miracle to also already be conformal.
These arguments of robustness of finiteness and speciality of conformality
should convince us that the two things are very different and that
our UV-finite theories are just a first step on the way towards full
conformality at the quantum level. In finite theories, a divergent part
of the effective action disappears, but the finite terms generically
do not enjoy conformal symmetry. They have to be much more constrained
to be called conformal effective actions on the quantum level.

Thus, what is the speciality on the level of finite terms behind the
UV-finite theories, theories without divergences, theories without
beta functions? There is some physically very interesting feature
of these theories---there is no running of couplings in these theories,
the couplings are always like bare ones. We do not need to do any
infinite renormalization of couplings. However, in the theories (typically
finite), we have different energy dimensions of couplings. They could
be both dimensionless and dimensionful. The lack of RG running is
quite special compared to other QFT models, and we are here at a fixed point of
RG. However, what is the impact of these conditions on the finite
terms in the effective action? In addition, here comes the basic information
about ingredients of infinite renormalization in QFT. If there is a running,
then we have to use some renormalization conditions and fix some couplings'
values at some physical scale $\mu$. For the construction of the
effective action, this means that we have at our disposal one more
dimensionful parameter---this scale $\mu$. If we do not have RG flow, we
do not have this complicacy.

Now, how do we construct various terms of the effective action (finite
terms)? We have to compensate the dimensions of various operators
(constructed with curvatures and covariant derivatives) by the dimensions
of other parameters present in the quantum theory. For example, due
to the RG flow, we need to include terms of the type $\log\square$ or
even more non-local terms like $\square^{-1}$ in the effective action.
As the parameters for our use, we have couplings in front of all operators and
here in finite theories we do not have a scale $\mu$, because of finiteness.
Is this very constraining? The~answer is no. Still, we can construct
operators of the type ${\cal R}{\cal F}_{2}(\square){\cal R}$ or
${\cal R}^{2}{\cal F}_{2}(\square){\cal R}$ and higher in curvature
too, by compensating the dimension of various operators with ratios of various couplings
of various dimensions. This is a general situation in higher derivative
theories or in non-local theories, where there is a scale of non-locality
at our disposal to construct such dimensionally viable terms in the
effective action. We conclude that a lot of different terms can still
appear in the effective action (in~finite terms) and UV-finiteness
here does not constrain them at all. This is the situation where we
have both dimensionful and dimensionless couplings. It is to be emphasized
that the operator additions that we had in the examples considered
above (${\cal R}^{5}$ and ${\cal R}$ or similar ones) come with
front coefficients, which are dimensionful in $d=4$. Moreover, the
front coefficients in the main action Eqn. (\ref{eq: 4daction}) are also
such. Thus, for these higher derivative UV-finite theories, we are precisely in
the situation described above.

However, let us imagine now that one-loop UV-finiteness is obtained in QFT models,
where there are only dimensionless parameters. How then we can construct
terms in the effective action (finite pieces)? We do not have $\mu$
at our disposal, and we do not have any other dimensionful parameter at the
quantum level because the classical couplings of the theory are all dimensionless.
Thus, we can really construct only very few operators 
in the effective action---only those that have the energy dimension
equal to the dimension of the spacetime, really very few. We know
what this would mean on the classical level: only scale-invariant
classical actions are acceptable (a first step towards conformality).
In addition,~the same would be repeated for the effective action at one-loop---the same construction, but maybe with different coefficients, but
always finite number of terms. Comparing the two actions (the original
classical and the quantum effective), we see that here we have only a
finite renormalizations of couplings, of dimensionless not RG-running
coupling constants. This is really constraining an effective action
very strongly. For example, for gauge theory in $d=4$, we could have
only in the one-loop effective action term of the form $\frac{1}{g'^{2}}F^{2}$,
if the theory is one-loop UV-finite, while in the classical action
we started with, a term precisely of the form $\frac{1}{g_{0}^{2}}F^{2}$, where $F$ is the gauge field strength.

There is a known example of such finite quantum theory (and not only
at one loop, but to all loop orders, and also non-perturbatively) called
an ${\cal N}=4$ SYM. However, at the one-loop level, there are many more
examples of theories scale-invariant classically, which are also UV-finite,
so their effective action at one-loop is very constrained and simple.
To mention one candidate here: simple bosonic gauge theory (YM) with precisely
tuned number of charged massless fermions. It is done as a textbook exercise
in QFT to find a condition for one-loop beta function of the gauge coupling to vanish in
such model, and the solution is simple, and the resulting theory is
one-loop UV-finite and one-loop conformal \cite{hollowood}. We remind
readers here that the scale-invariance (so no dimensionful parameters and
no mass or energy scales) of the action is a necessary condition for
classical conformal invariance of such theory, so on a tree-level. Now, we also understand
how this works at the loop levels, if the theory is very special like
UV-finite, so not possessing any divergences. Since we want to have
a conformal symmetry preserved also on the quantum level, then it seems
that the necessary condition is to have it also on the classical
level in the first place.

For the moment, here we have considered theories without conformal
compensators (dilatons). However, if we include it in the construction
of the action, then we may not have any dimensionful coupling or
parameter in the theory. This would be good news, since this is the first
step towards conformality. However, then during the construction of the
effective action, we can use various combinations of operators containing
this dilaton field, like in the denominators, in form-factors, in logarithms, etc.
We remind readers here that the dilaton carries an energy dimension. Thus, again in this situation,
the form of the effective action is not very constrained because how
can the dilaton which plays the role of the energy compensator be constrained?
We are back to the same case of a theory with dimensionful couplings!

We see that we can obtain conformality on the quantum level, only if we
start with a classically scale-invariant theory and without dilaton
(or when the dilaton field completely decouples dynamically). Only
when we have fine-tuning to get UV-finite theories at loop levels can we have this property of conformality preserved at the quantum
level. If we have terms with higher derivatives, then generically we
do not deal with conformal theories at a one-loop level, despite the fact that we may
have UV-finite theories quite easily (with a small level of fine-tuning
of only few parameters in the action) because still we have many dimensionful
couplings and parameters in the theory. Hence, the one-loop effective
action in such theories is not very constrained and this is not a feature exhibited
by conformality. Only for theories already scale-invariant on the
classical level, UV-finiteness on the quantum loop level implies conformality
on these loop levels. In addition, then the effective action is really very
constrained by the power of conformal symmetry there.

It seems that only matter theories with a small number of derivatives (like two)
can be conformally invariant at the one-loop level. The examples mentioned
above support this claim and in $d=4$ we know all classically conformally
invariant actions for all fields with a spin $s\leqslant2$. To have scale-invariant (or conformal) theories
on the classical level (tree-level), a two-derivative
massless scalar field theory is sufficient, but it can be also with a quartic potential $\phi^4$ (like the Higgs potential)
and when we want to couple it to curved geometry, we should also add a term of non-minimal
conformal coupling $\frac{1}{6}R\phi^2$. Fermions should be massless and without potential (so no Fermi interactions).
The standard two-derivative YM action $F^2$ is enough to get the classical conformality too.
Finally, we can couple all these matter species in the gauge invariant way using only dimensionless couplings
(Yukawa couplings are possible).
The total gauge
theory can be made one-loop conformal by the construction described
above (to add fermions in the precise number). To make it two-loop conformal, one also needs
to add scalars. In addition, from this level, one discovers that the spectrum
is basically the same like in ${\cal N}=4$ SYM theory, and since there
interactions are also highly constrained (like the spectrum is), then~only the (kinematical) spectrum is enough and it decides about finiteness and conformality
to all loop levels. Actually, in ${\cal N}=4$ SYM theory, the conformal
symmetry is preserved on the quantum level by the token that the conformal
current appears in the super-algebra of the theory, hence it must
be conserved perturbatively and also non-perturbatively, and by both
classical and quantum dynamics. The condition of conservation of the
conformal current is equivalent to the vanishing of conformal anomaly,
hence, in this theory, we have conformality all the time.

However, the situation with gravity is not so fortunate in $d=4$, since only $C^{2}$
gravitational action is classically conformally invariant in $d=4$.
We can also add $R^{2}$ term, since it is classically scale-invariant,
but this term is not conformally invariant though. However, even with
this arbitrary addition, it was found that, in pure gravitational theory,
we are not able to cancel
the beta functions for all gravitational couplings at one-loop, so
the theory is not one-loop conformal. Based on Eqn. \eqref{eq: divpart},
we need to cancel all four structures for quantum one-loop conformality. This is clearly improbable, since,
after removing the overall rescaling of the schematic action $R^2+C^2$,
we have only one dimensionless ratio parameter and four beta functions
to kill. However, when some matter multiplets
are added (motivated by supergravity), within the procedure
similar to the one described above for the ${\cal N}=4$ SYM, then,~according to Fradkin and
Tseytlin, it can be made UV-finite, so also one-loop conformal because then such theory
 contains only dimensionless coupling constants. Maybe this can
be extended to higher loops and non-perturbatively also. However, then,
which supergravity theory is it? What is its symmetry algebra? Is
there a place in the super-algebra for local conformal current?
Will it be also covariantly conserved whether the conformal symmetry
 can be understood on the
level of this super-algebra, like in the case of ${\cal N}=4$ SYM
theory? We can leave these questions aside, since this is a domain
of higher derivative conformal supergravity. We only discuss the Fradkin--Tseytlin
conformal supergravity briefly in Section \ref{sec:Conformal-Supergravity}.

Instead, here we propose a gravitational theory without matter, which
still may be one-loop conformal, but it must be defined in a dimension
higher than four. Let us be in $d=6$-dimensional spacetime and then
consider the following schematic gravitational action
\begin{equation}
S_{{\rm grav},6D}=\!\int\!d^{6}x\sqrt{|g|}\left(a{\cal R}\square{\cal R}+b{\cal R}^{3}\right)\,.\label{eq: 6daction}
\end{equation}

This action, as it stands, is classically scale-invariant (it may contain
not only Weyl tensor here, but~other curvature tensors as well) because, in $d=6$,
the coefficients $a$ and $b$ are dimensionless. It does not need
to be here an example of classically conformal action, since only scale-invariance
of the classical action is a prerequisite for quantum conformal invariance.
In $d=6$, this is not a super-renormalizable theory, but merely a
renormalizable gravitational one. However, let us concentrate on the
one-loop level only. The divergences are there, but they can be cancelled
in DR by adjusting the coefficients $b$ of cubic killers schematically
denoted by the terms of the type ${\cal R}^{3}$ above. The beta functions
for divergent covariant terms of the forms ${\cal R}\square{\cal R},\,{\cal R}^{3}$
depend quadratically on the $b$ coefficients. However, maybe it
is still possible to solve for them, get real values as solutions
and adjust them in such a way that the theory at one-loop becomes
finite. Then, this is an example of one-loop quantum conformally invariant
gravitational theory in $d=6$ spacetime dimensions. For a similar situation,
we may have it in eight-dimensional spacetime with the similar action
\begin{equation}
S_{{\rm grav},8D}=\!\int\!d^{8}x\sqrt{|g|}\left(a{\cal R}\square^{2}{\cal R}+b\left({\cal R}^{3},\nabla^{2}\right)+c{\cal R}^{4}\right)\,.\label{eq: 8daction}
\end{equation}

Here, all the beta functions are linear in the dimensionless parameters
$c$. However, naively, the number of the beta functions to cancel is the same
as the number of all terms in the above action, and this number is
simply greater than the number of possible mutually irreducible terms of the type ${\cal R}^{4}$.
Hence, it is quite unlikely to find non-zero solutions for the algebraic
system of equations for $a$, $b$ and $c$ coefficients. The dependence
on $b$ coefficients is quadratic and on $a$ even more complicated,
expressed through some rational functions. However, then maybe also this
system can be solved for the $c$ coefficients to get one-loop UV-finiteness
and hence one-loop conformality.

Let us here describe the general philosophy behind the construction
of minimal conformal theories in any dimension $d$. In a general dimension, scalars
and fermions (spin $\frac{1}{2}$ fields) still have to be described by two-derivative
actions to be classically scale-invariant. This means that also their conformal
dimensions depend on $d$. However, the gauge field potentials always have
the energy dimension one (because of the structure of the covariant derivative). Hence,
the construction of scale-invariant classical actions for gauge fields is more involved.
Basically, in a general dimension, we have to include all terms that are constructed in a covariant way
with gauge-covariant field strengths and with covariant derivatives acting on them, with
front coefficients that are dimensionless in given dimension $d$. We~couple all these fields by using only
dimensionless coefficients.
 The requirement of scale-invariance
is the prerequisite for quantum conformality. Of course, in this way, in the gauge sector, we
will inevitably have a dynamics described by HD theories. The conformality on the quantum level (recursively at
one-loop, two-loop level etc.) we gain by adjusting some values of the dimensionless coefficients
present in the action. With some level of fine-tuning we can reach theories, which will be UV-finite
and hence also conformal. Probably, the addition of supersymmetries will also help in this~task.

The situation with gravitational theories is a bit similar, but there are also
some significant differences. As the basis, we again need to use all terms that are with precisely $d$
derivatives on the metric in general dimension $d$. Therefore, the starting point
is the action for first renormalizable gravitational theory. All coefficients are dimensionless and
we have classical scale-invariance. In~the action for $d>4$, typically we also have some killers.
If we hope for one-loop conformality, then we need to fine-tune their coefficients. The counting of free
parameters and the beta functions is the following. In beta functions, we need to take care of all terms, also
of terms which are total derivatives or become such after variation (like the famous GB terms). However, these terms
are not important as killers; they are not even spectators because they do not contribute at all to the effective
action (even not to finite pieces), since their general variations vanish. Hence, we have more beta functions to kill
than the operators that we can use as killers. Moreover, the last argument, referred to also earlier,
 decreases the number of killers by one because of overall rescaling of the action. This counting explains,
 why, in $d=4$, we would need to kill four beta functions with only one free coefficient. As~we see,
 unfortunately, it is quite improbable to kill all the necessary beta functions, and only in the highly
 symmetric cases can we hope for full cancellation. That is why the study of symmetries of the
 gravitational action is very important.

As a last remark here, we want to put a comment about conformal theories
with dilaton. Then, with the conformal compensator, any theory that
is UV-finite can be made explicitly conformally invariant. This trick with using compensators is known for any symmetry. As explained
above, finiteness is a necessary condition for getting a conformal
theory on the quantum level. Now, by adding a dilaton, we check that the
effective action (finite terms) transforms covariantly under conformal
transformations, and actually the spacetime integral of this action is a conformal
invariant. Technically speaking, we~have conformal symmetry there.
However, what are the physical consequences of such theory with a dilaton,
which appears in many places to cancel the energy dimension?

If the theory is conformally invariant (even in the technical way
using the dilaton field), then~we are entitled to use conformal transformations
of the theory and, in this way, for example, we can succeed with the
programme of desingularization of classical gravitational field solutions.
With dilaton, we can transfer all singular behaviour to an unobservable
conformal factor (dilaton field) in such a way that the (conformal)
geometry is completely regular. The non-regular behaviour is now shifted
to a conformal gauge-dependent quantity (dilaton field) and in the
spacetime all particles feel a regular conformal spacetime, and move
in it never reaching a tentative singularity. We emphasize that it
was important in this programme first to achieve a construction of
UV-finite gravitational theory. Then, using the trick with a dilaton,
we made the theory manifestly conformally invariant. This let us conformally
desingularize all previously singular solutions. For the theory, which
is only classically conformally invariant, we cannot succeed with
such a programme. The problem with consistency lies on the quantum level. Namely,
if some putative theory has non-vanishing beta functions, then we cannot
consistently introduce the dilaton field and in this way we can never
achieve conformality on the quantum level (even in the technical mode,
when we exploit the dilaton field). The advantage of our theory is
that we can desingularize full quantum gravitational solutions (that
is solutions for the EOM coming from the quantum effective action),
because we have the theory conformal on the quantum level. Instead,
in theories, only classically conformally-invariant (but with a conformal
anomaly on the quantum level), we can remove singularities only from
exact classical solutions of the theory. However, then one can argue
that quantum effects can bring back ``bad'' singular solutions and
destroy the resolution of singularities, which was achieved only on
the classical level. In our UV-finite gravitational theory, this is
not a problem (because we have a very good control over quantum effects),
and we have a theory showing that all solutions are singularity-free.

However, we believe that, unfortunately, in this disguise for the UV-finite
gravitational theories with dilaton field, conformal symmetry does
not constrain too strongly the physical observables. We~saw already
that, on the level of finite terms of the effective action, we may have
many (infinitely many, in principle) terms with dilaton fields. They
are not constrained at all, conversely to the case of quantum conformal
theories without dilatons. However, let us speak about more physical quantities,
like the $S$-matrix elements.

\section{Scattering Amplitudes in Scale-Invariant Theories at the Tree-Level\label{sec:scattering}}

We have the canonical energy dimensions of fields of various spins
and in general dimension $d$:
\begin{itemize}
\item scalar $[\phi]=E^{(d-2)/2}$, so $d_{s}=\frac{d-2}{2}$,
\item fermion (spin 1/2) $[\psi]=E^{(d-1)/2}$, so $d_{f}=\frac{d-1}{2}$,
\item vector $[A]=E^{1}$, so $d_{v}=1$ and
\item two-rank tensor $[h]=E^{0}$, so $d_{t}=0$.
\end{itemize}

These $d_i$ are equal to conformal dimensions used for construction
of scale-invariant actions with the fields in question in general dimension $d$,
as this was done in the previous section.
In~scale-invariant theory on-shell amplitudes of generalized scattering
processes, in which participates precisely $n_{s}$ scalars, $n_{f}$
fermions, $n_{v}$ vectors and $n_{t}$ tensors are denoted by
\begin{equation}
S\left(p_{1}\ldots p_{n_{s}},q_{1}\ldots q_{n_{f}},r_{1}\ldots r_{n_{v}},s_{1}\ldots s_{n_{t}},\epsilon_{1}\ldots\epsilon_{n_{f}},\kappa_{1}\ldots\kappa_{n_{v}},\lambda_{1}\ldots\lambda_{n_{t}}\right),\label{eq: ampl}
\end{equation}
where $p$, $q$, $r$, and $s$ are momenta of respectively scalars,
fermions, vectors and tensors participating. By $\epsilon$, $\kappa$,
and $\lambda$, we denote the polarization spinors, vectors and tensors.
(We must have $n_{f}$ to be an even number.) Since in the scale-invariant
theory there is no any dimensionful parameter (at the tree-level), we have
the following asymptotic scaling property of the above amplitude under
a uniform rescaling of all momenta: $p_{i}\to\alpha p_{i}$
\[
S\left(\alpha p_{1}\ldots\alpha p_{n_{s}},\alpha q_{1}\ldots\alpha q_{n_{f}},\alpha r_{1}\ldots\alpha r_{n_{v}},\alpha s_{1}\ldots\alpha s_{n_{t}},\epsilon_{1}\ldots\epsilon_{n_{f}},\kappa_{1}\ldots\kappa_{n_{v}},\lambda_{1}\ldots\lambda_{n_{t}}\right)=
\]
\[
\underset{\alpha\to+\infty}{=}\alpha^{d-n_{s}d_{s}-n_{f}d_{f}-n_{v}d_{v}-n_{t}d_{t}}\times
\]
\[
\times S\left(p_{1}\ldots p_{n_{s}},q_{1}\ldots q_{n_{f}},r_{1}\ldots r_{n_{v}},s_{1}\ldots s_{n_{t}},\epsilon_{1}\ldots\epsilon_{n_{f}},\kappa_{1}\ldots\kappa_{n_{v}},\lambda_{1}\ldots\lambda_{n_{t}}\right)+
\]
\begin{equation}
+O\left(\alpha^{d-n_{s}d_{s}-n_{f}d_{f}-n_{v}d_{v}-n_{t}d_{t}-1}\right)=\label{eq: scatamp}
\end{equation}
\[
=\alpha^{d-n_{s}d_{s}-n_{f}d_{f}-n_{v}d_{v}}\times
\]
\[
\times S\left(p_{1}\ldots p_{n_{s}},q_{1}\ldots q_{n_{f}},r_{1}\ldots r_{n_{v}},s_{1}\ldots s_{n_{t}},\epsilon_{1}\ldots\epsilon_{n_{f}},\kappa_{1}\ldots\kappa_{n_{v}},\lambda_{1}\ldots\lambda_{n_{t}}\right)+
\]
\[
+O\left(\alpha^{d-n_{s}d_{s}-n_{f}d_{f}-n_{v}d_{v}-1}\right)\,,
\]
where in the last equality we used the fact that, for tensors (metric field--gravitons), we have $d_t=0$.
This is a result based on dimensional analysis only. Moreover, this is an asymptotics
for the UV regime of the scattering amplitude. Of course, the coefficient
in front of this leading term in the UV asymptotics depends on the
particular theory. However, there are no more constraints even if the
theory has classical conformal invariance. The only constraint on
scattering amplitudes comes from scale-invariance, which is of course
a prerequisite for the latter. By asking for Green functions with
states, which carry energy-momentum in form of $p_{i}$ (so they carry
an energy scale), we explicitly break the conformal invariance of the
problem. We warn, however, that this is a breaking ``by solution'', not spontaneous. The
asymptotic states correspond here to solutions of linearized EOM because
asymptotically all fields are very weak and are there in the form of plane
waves solutions $e^{ip_{i}x}$ multiplied by proper polarizations tensors. These
are genuine asymptotic states in the scattering problem. These solutions,
as in- and out-states, break the conformal symmetry of the vacuum (or
the hidden conformal symmetry of the vacuum, if the last theory was
in the spontaneously broken phase). For the $S$-matrix, we analyze
Green functions with only such asymptotic states, so we cannot have full
conformal symmetry in non-vacuum solutions (plane waves).

Instead, if we consider
Green functions of vacuum only (like vacuum to vacuum transitions
on curved spacetime or on non-trivial non-flat gauge connection backgrounds),
then we should still expect the presence of the full conformal symmetry.
Conformality should not to be broken there to scale-invariance,
and the constraints on such Green functions should be the consequences
of the full conformal symmetry of the theory. Here,
the token of scale-invariance is the same as doing dimensional analysis. Finally,
we want to emphasize that on-shell scattering amplitudes are different
than $n$-point functions in any CFT. The latter are constrained strongly
by the conformal symmetry. The former depend on the on-shell conditions,
so the plane wave solutions necessarily enter, and~that is why conformal
symmetry is broken ``by solution'' and only scale-invariance remains
and constrains amplitudes. Additionally, we remark that the problems
with asymptotic states are the problems with field configurations
localized at spatial and temporal infinity, so very far away from
the center, where collisions and interactions take place. This long
distance means that they are related to IR configurations. It is, moreover,
not surprising that the breaking of conformal symmetry by the introduction
of in- and out- states is equivalent to the problem of IR (infrared)
singularities of the conformal theory.

Therefore, if the theory is tree-level (that is classically) conformally
invariant, then there is no further constraints on scattering amplitudes besides
those coming from dimensional analysis,
and the amplitudes take the general form described above. Of course, the coefficients
of the leading term in the UV are very specific, but none can conclude
that these coefficients must vanish. In general, even for conformal theories,
this does not occur. These facts are confirmed by the study of scattering
amplitudes in tree-level scale-invariant theories or even fully conformal
on the quantum level, like ${\cal N}=4$ SYM theory in $d=4$. Actually,
the fact that the theory is quantum conformal does not influence anything
on the tree-level scattering amplitudes, since the loops never appeared
there. The coefficients are very special, but they do not vanish as
we know for the cases of 6-gluon (6-particle) scattering in such theory. The result
for this scattering is non-zero at tree-level for just pure YM theory
(classically scale- and conformally invariant, but not on the quantum
level) and also for ${\cal N}=4$ SYM theory (fully quantum conformal).
Therefore, conformal symmetry does not trivialize scattering amplitudes
at the tree-level. They are sufficiently constrained by scale-invariance
of the tree-level action. We want to emphasize that, despite the fact that in
the consideration of scattering amplitudes we focus on the tree-level
here, the~same results can be immediately applied to the scattering
on the quantum level in conformal theories on the quantum level. This
means that the same conclusions apply to the full quantum scattering
amplitudes in ${\cal N}=4$ SYM theory and also to scattering in UV-finite
models of quantum gravity.

There could be some exceptions to the general UV asymptotics as presented
above in a big formula Eqn. (\ref{eq: scatamp}) for the scattering amplitudes.
However, they come because of some other particular conditions and are
not related to conformal symmetry. Typically, this says that the asymptotics
is softer in UV than the one in the general formula (the exponent
on $\alpha$ is smaller than the generic one). For example, for scattering
of 3 gluons the result is zero, but this is because of parity symmetry
of the amplitude (asymptotics smaller than the naive $\alpha^{1}$
in $d=4$, which is based on the formula Eqn. (\ref{eq: scatamp}) applied to the pure YM theory). Here,
$\alpha$ is proportional to the energy $E$, so we really speak about
energy-scaling of UV amplitudes. Another example is a scattering of
gravitons in $d=4$ in Stelle theory (with Einstein--Hilbert term),
where the amplitude goes like $\alpha^{2}$, not like expected from
the general formula $\alpha^{4}$. This is a consequence of the redefinition
theorem and the Gauss--Bonnet identity, which are true together only
in $d=4$ \cite{scattering}. The fact that in pure Stelle quadratic
theory in four dimensions
\begin{equation}
S_{{\rm grav},4D}=\!\int\!d^{4}x\sqrt{|g|}\left(aR^{2}+bR_{\mu\nu}^{2}\right),\label{eq: 4dgrav}
\end{equation}
there is no scattering at all on the tree-level is not a consequence
of scale-invariance of this theory. It~is similarly not a consequence
of classical conformal invariance the more particular fact, that~in Weyl
square gravity, there is no graviton scattering at the tree-level
in $d=4$. These facts are simply consequences of the speciality of
four dimensions because only there we have together field redefinition
and the GB identity at work. As we see, the conformal enhancement of
the symmetries of the classical action (to Weyl square theory possessing
conformal invariance) completely does not improve the UV behaviour
of scattering amplitudes in this class of theories.

To continue with a similar example, but in $d=6$, let us consider
a theory schematically written as
\begin{equation}
S_{{\rm grav},6D}=\!\int\!d^{6}x\sqrt{|g|}\left(a{\cal R}\square{\cal R}+b{\cal R}^{3}\right)\,.\label{eq: 6dgrav}
\end{equation}

In such theory, any tree-level scattering amplitude with an even number
of gravitons (bigger than 2) will go asymptotically in UV generically
like $\alpha^{6}$. Due to the Furry's theorem for quantum gravity
(which is motivated by parity symmetry preserved in such theory), we
decided not to consider amplitudes with an odd number of gravitons. A
classically conformally invariant action in this dimension is given
schematically by
\begin{equation}
S_{{\rm grav},6D}=\!\int\!d^{6}x\sqrt{|g|}c_{i}\left(C^{3}\right)_{i}\label{eq: 6dconf}
\end{equation}
(various contractions of three Weyl tensors; algebra of indices says that
there are only two such contraction possibilities). However, in the theory Eqn. \eqref{eq: 6dconf},
there is no propagator around flat spacetime, only~vertices and all
the Green functions are already 1PI diagrams (vertices) at the tree-level.
This signifies that, for computation of tree-level scattering amplitudes,
we do not need to consider Feynman diagrams here with propagators
and internal lines. All amplitudes seem to be read from the vertices
of the theory only. A naive, and moreover incorrect, application of the token
of conformal symmetry here could say that the scattering amplitudes with gravitons
at the tree-level should not contain any power of the energy $E$ of gravitons, regardless of
the actual number of them participating in the scattering. However, it would be very strange to say that all scattering
amplitudes with gravitons are zero, while the vertices are there non-vanishing
on flat spacetime, which is why again the conformal symmetry does not
constrain anything regarding the scattering amplitudes here. Actually, we cannot speak about on-shell perturbative
states here, so there is no $S$-matrix! (If the ``naive'' argumentation
is accepted that some supposed amplitudes cannot be proportional to
$\alpha^{6}$, then conformal symmetry would require all of them to
vanish; moreover, all vertices should vanish, so there would be no theory at all at the tree-level).

Instead, in this theory, we have non-trivial
classical Green functions with $n\geqslant3$ (due to vertices) even on flat
spacetime, but we do not have any perturbative scattering amplitudes, since
there is no propagation of gravitational perturbations. Here, we do not have at our
disposal the redefinition theorem, neither GB identity. What we have
instead is the fact that theory is higher than second in curvature,
so there are no well-defined perturbative asymptotic states. There
are no asymptotic states because we cannot define a propagator around
flat spacetime. Even some attempts with defining asymptotic states
of this theory around maximally symmetric backgrounds MSS (like de
Sitter and anti-de Sitter) will fail either because the theory is
higher than quadratic in Weyl tensor and again the propagator does
not exist on any conformal background, which MSS manifolds are example
of. Hence, we know that generic amplitudes in such tree-level (classically)
conformally invariant theory vanish, but again this is not a direct
consequence of conformality, but of other special circumstances. The similar conclusion
can be derived for any theory with the gravitational Lagrangian ${\cal R}^3$ or higher. The
same happens for gauge theories higher in field strengths than quadratic,
which are respectively conformal in higher dimensions.

It is now obvious that in such conformal theories on the classical
level there is no any problem with the unitarity bound, since the amplitudes
are very well-behaved in the UV regime. In the theory above Eqn. (\ref{eq: 6dconf}),
there are no amplitudes, in the theory $C^{2}$ in four dimensions
amplitudes could be defined, but they also vanish. In Fradkin--Tseytlin,
conformal supergravity based on the bosonic gravitational sector in
four-dimensional spacetime on the same $C^{2}$ term, the scattering
amplitudes of any number of gravitons have very mild UV-dependence. As we know,
the results in such theory are valid both on the classical tree-level,
as well as on the quantum level, because the theory is quantum conformal.
The channels with graviton exchange contribute nothing and we have
only Feynman diagrams with mediation of scalars and gluons of ${\cal N}=4$
SYM theory. It is well known that the scattering of gravitons due
to only virtual gluons and scalars gives rise to the amplitude, which
in the UV limit tend to a constant, which is related to the SYM gauge
coupling constant. In addition, of course, such a gauge coupling constant does
not put a unitarity bound in danger at all. Hence, we see that, in all
examples considered here of conformally invariant gravitational theories, the unitarity
bound is always satisfied.

The last remark is about the theory with dilaton (conformal compensator).
Then, the theory is without any dimensionful parameter, so it is at tree-level
scale-invariant. The amplitudes should follow the general formula
Eqn. (\ref{eq: scatamp}) presented above. However, the dilaton takes the
vacuum expectation value (v.e.v.). Thus, this means that in the vacuum
of the theory there is a new energy scale, or simply this vacuum is
not scale-invariant. We cannot have dilaton on external on-shell legs,
but its v.e.v. can be a new parameter, that amplitudes may depend upon.
Since the scale-invariance is broken even by the vacuum choice (dilaton
v.e.v.), then here the scale invariance is not broken only by solutions,
 and the derivation above for the general formula Eqn. (\ref{eq: scatamp}) does not hold. The
UV asymptotics of physical amplitudes is not constrained by scale-invariance
and it could be quite complicated. If the dilaton is frozen in its
v.e.v. and we cannot have it on external legs, then this is like a theory
with many mass scales, so there is no constraint on scattering amplitudes
besides the proper energy dimension, that~they should have (indeed,
this is what the dilaton checks). The dilaton can be also a virtual
dynamical excitation, but not the physical one. However, this still does
not reintroduce scale-invariance or the conformal invariance constraints
on the scattering amplitudes.

In conformal theories without dilaton (these are highly constrained
theories too), the scattering matrix elements can depend only in a
very precise way on the energy of colliding particles Eqn. \eqref{eq: scatamp}. This~way is
dictated by scale-invariance of the theory. Otherwise, this would
violate assumed conformality, which implies scale-invariance. This
is the case of one-loop conformal theories, where scattering amplitudes
at one-loop are the same (up to a finite renormalization of coupling
constants) as the tree-level ones. Thus, indeed, they are very simple
and all effects of quantum dynamics are washed away in the structure of these
amplitudes.

What about scattering of physical excitations in conformal theories
on the quantum level with dilatons? Looking at the effective actions,
which generate such amplitudes, we notice that they are not very constrained.
They are quite general and have many terms. Thus, what is the role of
conformal symmetries embodied here in the form of dilaton? We think,
that its effect is quite simple. Basically, it constrains the amplitudes
in such a way that they have the correct energy dimensions. No more constraints---only the constraint of overall energy dimensions, which we could
derive by ourselves easily earlier by dimensional analysis, without
the need to employ dilatons and conformal symmetry. This is because
the dilaton stays there, where it is needed to compensate for unbalanced
energy dimension for some operators. Like emphasized above, the true
power of conformal symmetry is to be seen, when the dilaton field dynamically
decouples. Moreover, for such scattering amplitudes of physical particles,
we cannot have a dilaton as external particle because, by using conformal
symmetry, we can always gauge it away. Thus, its presence on the external
legs is conformal gauge-dependent. In addition,~as a physical excitation, we
cannot allow it because it depends on a gauge. This concludes the
part about scattering amplitudes in conformal theories with dilatons.

The following part contains some new arguments and understanding of
the issues, which were discussed for a long time about finite terms,
${\cal N}=4$ SYM theory, effective actions, scattering amplitudes,
dilatons and conformal symmetry.

Before, we put two statements about the $S$-matrix in conformally invariant theories.
It could be possible to reconcile the two, despite the fact that naively they seem to be contradictory.
 The first claim was that the $S$-matrix
does not exist because one cannot speak about any asymptotic state.
The second one is that, when one does computation, there is a formula
for scattering amplitude, which is constrained only by scale-invariance.
These statements could be both true at the same time and below we explain how
this happens.

One must think operationally how one would check the statements above.
In both cases, one would have to compute the scattering amplitudes.
Imagine that we could do it, for example, with a computer. Then, there
is a unique well-defined result, but to do this we had to pick up
some form of the asymptotic states. In addition, this act of picking up solutions
breaks conformal invariance, while in the conformal theory in question there
should be no asymptotic states at all, if there is conformal invariance unbroken,
which means that it is unbroken also by solutions, or, in other words, how would one operationally prove that the $S$-matrix
does not exist? There~could not be any explicit proof of the calculation
even for some specific theory. Why?---because then one uses some asymptotic
states, so one has broken (maybe accidentally and unawarely) conformal
symmetry. Thus, this is a fact, which is not provable by any analytic computation.
It is a statement that is to be proven only by asking for the absence
of asymptotic states due to the symmetries of the theory.

Now, for our purposes, for energies above the Planck energy, we can
think of the phase in which conformal symmetry is in the unbroken
state. In addition, then use the second statement that one cannot speak about asymptotic
states and that there is no problem with a unitarity bound of the
$S$-matrix. There is no $S$-matrix there for $E>M_{\rm Pl}$! One can say that, if someone
tries to insist on computing the $S$-matrix in such a phase, then he
will break the conformal invariance by the introduction of asymptotic
states, so the symmetry is gone, but in the symmetric phase we surely
do not want this to happen.

One can ask, in the light of previous comments, what is the meaning
of this whole industry of computing amplitudes in ${\cal N}=4$ SYM
theory---because, if from the beginning the conformal symmetry is used,
then the answer should be trivial---there are no asymptotic states,
no scattering matrix. In our opinion, the situation is as follows.
The ${\cal N}=4$ SYM is indeed a very special, conformal theory, without any doubts. However,
when people insist on computing tree-level amplitudes, then they reduce
its speciality and the conformal symmetry is gone. In addition, they find amplitudes
like in any local two-derivative gauge theory with the fermionic and
scalar matter. We mean that the result for the structure of these
amplitudes is not constrained by conformal symmetry. Of course, for
a particular theory, the~coefficients in amplitudes are precisely
and uniquely defined; however, there is nothing of conformal symmetry
there. It is like a scattering in any scale-invariant gauge+fermions+scalars
system at the tree-level. The conformal symmetry of ${\cal N}=4$ SYM
does not kill any term in the amplitudes and likely all possible terms
in such a situation are generated. The coefficients in the amplitudes
are only special because the coefficients in the Lagrangian of ${\cal N}=4$
SYM are very special, but~they are not zero. The true power of the
conformal symmetry comes only because of the loop level. As~we know,
the total quantum effective action is there identical to the original
action with an only finite shift of the YM gauge coupling, there is no divergent
part, etc. In addition, to prove this by explicit one-loop computation is quite
tedious. However, knowing all tree-level results now in the industry
of amplitudes the results of quantum amplitudes could be also firmly
and easily predicted. We also note an interesting correspondence between
scattering amplitudes in conformal gravity based on $C^2$ action and
similar amplitudes in YM theories in $d=6$ \cite{Johansson:2017srf,Johansson:2018ues}.

If in the conformally invariant theory we cannot define $S$-matrix,
by construction, then all problems of non-unitary higher derivative
theories disappear because the question of unitarity can be only asked,
if we can speak about the $S$-matrix. Then, higher derivative quantum
conformal theories satisfy all the requirements of the consistent
theories for quantum gravity.

\section{Conformal Supergravity\label{sec:Conformal-Supergravity}}

In the quest for a conformal quantum gravity, we must note that the
first successful model was formulated in the framework of extended
Weyl square supergravity theory by Fradkin and Tseytlin in 1984
\cite{FradkinT1,FradkinT2,FradkinT3,FradkinT4,FradkinT5,nlsugra}.
The authors used ${\cal N}=4$ supergravity theory based on higher
derivative action coupled to two copies of ${\cal N}=4$ SYM theory.
However, they were not able to find full symmetries of the coupled
system and they stated that this system was with eight explicitly visible
supercharges. One may think that the possible theory with more supercharges
would put even stronger constraints, like an ${\cal N}=8$ E--H supergravity
puts on the dynamics in the gravitational supermultiplets there. The crucial
thing in the construction by Fradkin and Tseytlin was the usage of
$C^{2}$ action in four dimensions as the action in purely gravitational bosonic
sector. This is in distinction to the case of other supergravity models,
which were based on two-derivative dynamics given by E--H action. The
obvious advantage is that such theory is naturally, at a classical level,
conformally invariant, while the E--H supergravity is not. The latter
contains a dimensionful parameter---Planck's mass. In the former theory
(Weyl square gravity), the effective Newton's constant arises at the
solution level because all the solutions (except conformally flat
manifolds) break the conformal invariance. This is similar to the
situation in classically conformally invariant four-dimensional electrodynamics,
where the Coulomb potential solution comes with a scale and this solution
explicitly breaks conformal invariance of the theory. Since in $C^{2}$
gravity we have conformal invariance at the beginning, it is not difficult
to keep it also at the quantum level after coupling it to a sufficient
amount of matter, which is superconformal by itself (but on the flat
background).

This conformal supergravity theory comes in four dimensions without
any mass scale. This is not surprising since scale invariance 
(absence of mass scales) is a first condition towards conformal
invariance on the classical level. It seems that only in four dimensions we are very
lucky that we can study scattering processes in conformal theory.
Moreover, with Fradkin--Tseytlin theory, in opposition to quantum conformal gravity
presented in Section \ref{sec:CQG}, we have the luxurious situation
that the couplings of the theory are dimensionless, so the theory is
scale-invariant also on the classical level. In~this very interesting
theory, the symmetry algebra is so big that the considerations about
the effective action from the Section \ref{sec:Conformal-symmetry-of}
are all true. Basically, the quantum effective action at any loop order coincides
in the structure with the classical action (and we could have only finite renormalizations of couplings).
This means that this theory is very simple, very beautiful and very
constrained on the quantum level.

Fradkin and Tseytlin explicitly computed and showed that all beta
functions in such theory vanish to perturbative and also to non-perturbative
level as the result of very high symmetry of the theory and precisely
chosen spectrum of matter fields coupled to conformal supergravity.
In their approach, the conformal symmetry is another gauge symmetry
of gravitation, together with diffeomorphism symmetry and with $SU(N)$
gauge symmetries of the matter sector. All these symmetries are local
as they should be in a consistent quantum gravity theory. The only apparent
problem as the proponents of this revolutionary theory had noticed,
was that this theory treated perturbatively to compute scattering
matrix elements gives naively non-unitary results. However, the resolution
to this puzzle can be also found above in this article. First, as emphasized in Section
\ref{sec:Story-of-infinities}, pure monomial supergravity theory based on $C^2$ action
in the gravitational sector exhibits a propagator with only one pole, but with multiplicity 2, before a splitting.
 This is a pole in a spin-2 sector and without a mass.  The ghosts appear
 when one insists on doing a  splitting 
and a resolution of this double pole into single poles. 

 On the other hand, since the quantum theory of F--T supergravity is conformal
 (all quantum beta functions vanish), then strictly speaking and according to
 arguments presented in the previous section, we should not speak about
 scattering processes in this conformal theory at all. Thus, all the results obtained by
 insisting on using asymptotic states bring with them the artifacts of breaking
 the conformal symmetry. This could explain the first initial observations
 made by the authors about apparent violation of unitarity in scattering processes.
 Additionally, we should remind readers that precisely in $d=4$ dimensions we do not have any scattering
 in the graviton sector, since there we could use the redefinition theorem and GB identity, like this was done already in Section \ref{sec:scattering} for a general four-dimensional
  Stelle's quadratic theory. We think that, if supersymmetry is powerful enough, then
  this result can be also transferred to the matter sector, or could
  constrain the mutual interactions between matter and gravitation. Furthermore,
  if the supersymmetry in the local version is powerful enough, then it could transmit
  the preservation of the unitarity from the gauge sector (where it is clearly there for YM gauge fields)
  to the graviton sector. Then, the issue with unitarity in this theory would be completely explained and we could view
  this model as a fully viable candidate for conformal quantum gravity.

There is one additional thing that could be expected from the highly
supersymmetric and conformal theory of supergravitation. This is the
question of grand unification of gravity and matter fundamental interactions.
Unfortunately, in ${\cal N}=4$ supergravity as introduced by Fradkin
and Tseytlin, the number of supercharges and related supersymmetries
is too small to relate matter and gravity sector in a very tight way,
which would constitute a true unification of all interactions. Instead, such strong
tights are known to exist for ${\cal N}>4$ supergravities based on E--H action,
where the matter sector and interactions in it are strictly related
to what happens in gravity. In~the language of algebraic classification
of particle representations, the supermultiplet, which contains gravity
in such higher supersymmetric theories contains also parts (or whole,
like in the ${\cal N}=8$ supergravity case) of the coupled supersymmetric
matter sector supermultiplets. It is known that in $d=4$ the number of conformal
supercharges is constrained and cannot exceed 16,
even for higher derivative supergravitational theories \cite{Cordova:2016emh}.
In the theory by Fradkin and Tseytlin, we have eight conformal supercharges.
 If somehow miraculously the ${\cal N}=8$ E--H supergravity is a UV-finite
theory, then we should expect there this maximal number of 16 superconformal supercharges
(maybe not all of them will be explicit, some maybe realized as hidden
symmetries of this theory). This would suggest that we could also try to extend the F--T supergravity
for a bigger number of conformal supercharges, like for the situation in ${\cal N}=8$ theory. However,
such extension is at the moment beyond our reach.
We~see that, in principle, there is a space
for further development of the model originally constructed by Fradkin
and Tseytlin.

Finally, we can mention that, in the Fradkin--Tseytlin conformal supergravity,
there are two dimensionless coupling constants: the coupling in front
of the Weyl square gravitational term $\alpha_{C^{2}}$ and the coupling
in front of gauge fields $g_{{\rm YM}}$. In principle, they are completely
independent constants. In a theory where the unification is successful,
 there should be a very strict algebraic relation between these
two couplings. We could expect such relation as a consequence of supersymmetry
transformations mixing matter and gravitational degrees of freedom.
In the bigger supermultiplet, we should expect both matter and gravitational
fields, and this is also another expression of the unification idea: to
put all interacting fields in the same multiplets of the same symmetry
group (this may not be a bosonic symmetry, but could be a fermionic
one, like a supersymmetry is). However, this is not a situation in Fradkin--Tseytlin
supergravity. We expect that a higher conformal supergravity should
be able to give clues for such unification of couplings, and matter
with gravitation in the same symmetry supermultiplets.

\section{Conclusions}

In this review, we showed explicitly what were the main problems of
quantum gravity formulated in the language of quantum field theory.
We also presented a construction of a theory that overcomes all
these problems. The final theory is without divergences in the UV
regime and without classical singularities for small distances (coincident
limits). It was essential to employ conformal symmetry in the local
version in the gravitational framework to succeed with these goals.
The conformal symmetry of the final theory is present both on the
classical and on the quantum level. Technically, we see it manifest, when we
use a dilaton field. The theory presents a very controllable behaviour
on the quantum level, which is equivalent to vanishing of conformal
anomaly. The version of a theory, in which all couplings are dimensionless,
was also discussed. Furthermore, we mentioned Fradkin--Tseytlin theory with increased
level of supersymmetries in a local version. We presented some new understanding related
to the form of the effective actions and scattering amplitudes in classical
and quantum conformal field theories. We advocated the virtues and specialities
of the known quantum CFT models. It is remarkable that quantum gravity is now
among them.

It is obvious that, in the real world, conformal symmetry is broken.
However, this does not mean that the formulation of the fundamental theory
does not enjoy conformal symmetry on the theory level. It~is not known
if in the real world the conformal symmetry is broken by the vacuum
choice (v.e.v. of the dilaton) or by solutions, which necessarily break
it, or it is just an approximate symmetry and it was never exact, or maybe
 there exists and is realized a still different method of its breaking. In
this former case, the world can experience scale-dependence and RG flows
due to some operator added to the CFT, which describes an UV fixed point.
 This perfect CFT with gravitational interactions is a starting point for the conformal perturbation theory.
To push forward such a programme, all the CFT data should be known about
the operators present in the CFT. We will compute in future the anomalous
scaling dimensions of primary (and also of higher generation) operators in this
gravitational CFT model. Another point is about the detailed study
of SSB of conformal symmetry and the choice of the profile for the
dilaton field. We just remark that, contrary to the case of scalars
in particle physics, which in the vacuum state can only take constant
values; here, the dilaton field can take some spacetime dependent profiles.
Besides the issue of breaking of conformal symmetry, we shall also
investigate in the future the role of conformal symmetry for the
form of exact gravitational solutions. We believe that conformal symmetry
constrains them too in a similar way like it constrained effective
actions and scattering amplitudes. Moreover, this last issue is
related to the possible ways of breaking of conformal symmetry
in the real world.

Having constructed the basis and the frame for quantum conformal gravitational
interactions, we~can now start studies of the issues related to breaking
of conformal symmetry and its impact on the gravitational theory.
The research on conformal gravity is quite popular and its various
aspects are already being investigated. We refer the interested reader
for the literature in Refs. \cite{hooft1,hooft2,hooft3,hooft4,litimcft2,mannheim1,mannheim2,mannheim3}.
It is also of great interest to focus on a possibility of experimental
or observational verification of the conformal gravitational theory.
Some studies in this direction were already performed in \cite{expjizba,exp1,exp2}.
We believe that, due to the high symmetry of the theory and the mathematical
beauty present in its construction, the verification that this is
the correct theoretical framework for quantum gravitational interactions
will come soon.

\vspace{6pt}
\acknowledgments{L.R. would like to thank prof. Leonardo Modesto at SUSTech, in Shenzhen, China for warm hospitality and the invitation to the conference ``International Conference on Quantum Gravity''.}

\funding{This research received no external funding.}
\conflictsofinterest{The author declares no conflict of interest.} 

\reftitle{References}


\begin{thebibliography}{999}
\bibitem{confan} Asorey, M.; Gorbar, E.V.; Shapiro, I.L. Universality and ambiguities of the conformal anomaly. \emph{Class. Quantum Gravity} \textbf{2003}, {\em 21}, 163.

\bibitem{confan2} Capper, D.M.; Duff, M.J. Trace anomalies in dimensional regularization.
\emph{Nuovo Cimento  A} \textbf{1974}, {\em 23}, 173--183.

\bibitem{confan3} Capper, D.M.; Duff, M.J. Conformal Anomalies and the Renormalizability Problem in Quantum Gravity. \emph{Phys. Lett. A} \textbf{1975}, {\em 53}, 361--362.

\bibitem{confreview}
Duff, M.J. Twenty years of the Weyl anomaly. \emph{Class. Quantum Gravity} \textbf{1994}, {\em 11}, 1387.

\bibitem{Kallosh} 
Kallosh, R.; Linde, A.D.; Linde, D.A.; Susskind, L. Gravity and global symmetries. \emph{Phys. Rev. D} \textbf{1995}, {\em 52},~912--935. 


\bibitem{Banks} 
Banks, T.; Seiberg, N. Symmetries and Strings in Field Theory and Gravity. \emph{Phys. Rev. D} \textbf{2011}, {\em 83}, 084019.


\bibitem{hove} t Hooft, G.; Veltman, M.J.G. One loop divergencies in the theory of gravitation. \emph{Ann. Inst. H. Poincare Phys. Theor. A} \textbf{1974}, {\em 20}, 69--94.

\bibitem{nieuwe}Van Nieuwenhuizen, P.; Wu, C.C. On Integral Relations for Invariants Constructed from Three Riemann Tensors and their Applications in Quantum Gravity. \emph{J. Math. Phys.} \textbf{1977}, {\em 18}, 182--186.

\bibitem{gosa1} Goroff, M.H.; Sagnotti, A. The Ultraviolet Divergences Of Gravity Theories. In Proceedings of the Fourth Marcel Grossmann Meeting on General Relativity,  Rome, Italy, 17--21 June 1985; UCB-PTH-85-42. 

\bibitem{gosa2} Goroff, M.H.; Sagnotti, A. The Ultraviolet Behavior of Einstein Gravity. \emph{Nucl. Phys. B} \textbf{1986}, {\em 266},~709--736.

\bibitem{gosa3} Goroff, M.H.; Sagnotti, A. Quantum Gravity At Two Loops. \emph{Phys. Lett. B} \textbf{1985}, {\em 160B},~81--86.

\bibitem{smax1}
Bern, Z.; Carrasco, J.J.M.; Johansson, H. Progress on Ultraviolet Finiteness of Supergravity. \emph{Subnucl. Ser.} \textbf{2011}, {\em 46}, 251--276.

\bibitem{smax2}
Bern, Z.; Carrasco, J.J.; Dixon, L.J.; Johansson, H.; Kosower, D.A.; Roiban, R. Three-Loop Superfiniteness of N = 8 Supergravity. 
\emph{Phys. Rev. Lett.} \textbf{2007}, {\em 98}, 161303.

\bibitem{smax3}
Bern, Z.; Dixon, L.J.; Roiban, R. Is N = 8 supergravity ultraviolet finite? \emph{Phys. Lett. B} \textbf{2007}, {\em 644}, 265--271.

\bibitem{smax4}
Banks, T. Arguments Against a Finite N = 8 Supergravity. \emph{arXiv}  \textbf{2012}, arXiv:1205.5768.
\bibitem{stelle1} Stelle, K.S. Renormalization of Higher Derivative Quantum Gravity. \emph{Phys. Rev. D} \textbf{1977}, {\em 16}, 953.

\bibitem{stelle2} Stelle, K.S. Classical Gravity with Higher Derivatives. \emph{Gen. Relat. Gravit.} \textbf{1978}, {\em 9}, 353--371.

\bibitem{FT1} 
Fradkin, E.S.; Tseytlin, A.A. Renormalizable Asymptotically Free Quantum Theory of Gravity. \emph{Phys. Lett. B} \textbf{1981}, {\em 104B}, 377--381.

\bibitem{FT2} 
Fradkin, E.S.; Tseytlin, A.A. Renormalizable asymptotically free quantum theory of gravity. \emph{Nucl. Phys. B} {\textbf{1982}}, {\em 201}, 469--491.

\bibitem{FradkinT3} Fradkin, E.S.; Tseytlin, A.A. Conformal Anomaly in Weyl Theory and Anomaly Free Superconformal Theories. \emph{Phys. Lett. B} \textbf{1984}, {\em 134B}, 187--193.



\bibitem{asorey} Asorey, M.; Lopez, J.L.; Shapiro, I.L. Some remarks on high derivative quantum gravity. \emph{Int. J. Mod. Phys. A} \textbf{1997}, {\em 12}, 5711--5734. 

\bibitem{Kluson}
Kluson, J.; Oksanen, M.; Tureanu, A. Hamiltonian analysis of curvature-squared gravity with or without conformal invariance. 
\emph{Phys. Rev. D} \textbf{2014}, {\em 89}, 064043.


\bibitem{Riegert}
Riegert, R.J. The Particle Content Of Linearized Conformal Gravity. \emph{Phys. Lett. A} \textbf{1984}, {\em 105}, 110--112.


\bibitem{T1} 
Tomboulis, E. 1/N Expansion and Renormalization in Quantum Gravity. 
\emph{Phys. Lett. B}  \textbf{1977}, {\em 70B}, 361--364.

\bibitem{T2} 
Tomboulis, E. Renormalizability and Asymptotic Freedom in Quantum Gravity. \emph{Phys. Lett. B} \textbf{1980}, {\em 97B},~77--80.

\bibitem{Kaku} 
Kaku, M. Strong Coupling Approach to the Quantization of Conformal Gravity.
\emph{Phys. Rev. D} \textbf{1983}, {\em 27}, 2819.
\bibitem{Bender1} 
Bender, C.M.; Mannheim, P.D. Exactly solvable PT-symmetric Hamiltonian having no Hermitian counterpart. 
\emph{Phys. Rev. D} \textbf{2008}, {\em 78}, 025022. 


\bibitem{Bender2} 
Bender, C.M.; Mannheim, P.D. No-ghost theorem for the fourth-order derivative Pais-Uhlenbeck oscillator model. \emph{Phys. Rev. Lett.} \textbf{2008}, {\em 100}, 110402. 

\bibitem{Tkach} 
Tkach, V.I. Towards Ghost-Free Gravity and Standard Model. 
\emph{Mod. Phys. Lett. A} \textbf{2012}, {\em 27}, 1250131. 

\bibitem{Smilga} 
Smilga, A.V. Supersymmetric field theory with benign ghosts. 
\emph{J. Phys. A} \textbf{2014}, {\em 47}, 052001. 

\bibitem{tomb1} Tomboulis, E.T. Superrenormalizable gauge and gravitational theories.  \emph{arXiv} \textbf{1997}, arXiv:hep-th/9702146.

\bibitem{tomb2} Tomboulis, E.T. Renormalization and unitarity in higher derivative and nonlocal gravity theories. \emph{Mod. Phys. Lett. A} \textbf{2015}, {\em 30}, 1540005.

\bibitem{tomb3} Tomboulis, E.T. Nonlocal and quasilocal field theories. \emph{Phys. Rev. D} \textbf{2015}, {\em 92}, 125037.

\bibitem{unita} Briscese, F.; Modesto, L. Cutkosky rules and perturbative unitarity in Euclidean nonlocal quantum field theories. \emph{arXiv} \textbf{2018}, arXiv:1803.08827.

\bibitem{unita2} Christodoulou, M.; Modesto, L. Reflection positivity in nonlocal gravity.  \emph{arXiv} \textbf{2018}, arXiv:1803.08843.
\bibitem{fakeunit} Asorey, M.; Rachwal, L.; Shapiro, I.L. Unitary Issues in Some Higher Derivative Field Theories. \emph{Galaxies} \textbf{2018}, {\em 6}, 23. 



\bibitem{scattering} Dona, P.; Giaccari, S.; Modesto, L.; Rachwal, L.; Zhu, Y. Scattering amplitudes in super-renormalizable gravity. \emph{J. High Energy Phys.}  \textbf{2015}, {\em 2015}, 38. 

\bibitem{fingauge} Modesto, L.; Piva, M.; Rachwal, L. Finite quantum gauge theories. \emph{Phys. Rev. D} \textbf{2016}, {\em 94}, 025021. 

\bibitem{exactsol} Modesto, L.; Rachwal, L. Exact solutions and spacetime singularities in nonlocal gravity. \emph{J. High Energy Phys.} \textbf{2015}, {\em 2015}, 173. 

\bibitem{entanglement} Giaccari, S.; Modesto, L.; Rachwal, L.; Zhu, Y. Finite Entanglement Entropy of Black Holes. \emph{Eur. Phys. J. C} \textbf{2018}, {\em 78}, 459. 

\bibitem{starobinsky} Koshelev, A.S.; Modesto, L.; Rachwal, L.; Starobinsky, A.A. Occurrence of exact $R^2$ inflation in non-local UV-complete gravity. \emph{J. High Energy Phys.} \textbf{2016}, {\em 2016}, 67.

\bibitem{confgr} Modesto, L.; Rachwal, L. Finite Conformal Quantum Gravity and Nonsingular Spacetimes. \emph{arXiv} \textbf{2016}, arXiv:1605.04173.

\bibitem{spcompl} Bambi, C.; Modesto, L.; Rachwal, L. Spacetime completeness of non-singular black holes in conformal gravity. \emph{J. Cosmol. Astropart. Phys.} \textbf{2017}, {\em 1705}, 3. 

\bibitem{evaporation} Bambi, C.; Modesto, L.; Porey, S.; Rachwal, L. Formation and evaporation of an electrically charged black hole in conformal gravity. \emph{Eur. Phys. J. C} \textbf{2018}, {\em 78}, 116. 

\bibitem{universality} Modesto, L.; Rachwal, L. Universally finite gravitational and gauge theories. Nucl. Phys. B 2015, {\em 900}, 147. 

\bibitem{superrenfin} Modesto, L.; Rachwal, L. Super-renormalizable and finite gravitational theories. \emph{Nucl. Phys. B} \textbf{2014}, {\em 889}, 228--248. 

\bibitem{evaporationconf} 
  C.~Bambi, L.~Modesto, S.~Porey and L.~Rachwal, Black hole evaporation in conformal gravity.
\emph{J. Cosmol. Astropart. Phys.} \textbf{2017}, {\em 1709}, 33.

\bibitem{mssfin} Modesto, L.; Rachwal, L. Finite conformal quantum gravity and spacetime singularities. \emph{J. Phys. Conf. Ser.} \textbf{2017}, {\em 942}, 012015. 

\bibitem{finmss} Koshelev, A.S.; Kumar, K.S.; Modesto, L.; Rachwal, L. Finite Quantum Gravity in (A)dS. \emph{arXiv} \textbf{2017}, arXiv:1710.07759.




\bibitem{RGsuperren} Modesto, L.; Rachwal, L.; Shapiro, I.L. Renormalization group in super-renormalizable quantum gravity. \emph{arXiv} \textbf{2017} arXiv:1704.03988.

\bibitem{finproc} Modesto, L.; Rachwal, L. \emph{Finite Quantum Gravity in Four and Extra Dimensions};  World Scientific Publishing Co.: Singapore, 2017; doi:10.1142/9789813226609\_0084.
\bibitem{singproc} Modesto, L.; Rachwal, L. \emph{Spacetime Singularities in Nonlocal Gravity};  World Scientific Publishing Co.: Singapore, 2017; doi:10.1142/9789813226609\_0305.

\bibitem{review} Modesto, L.; Rachwal, L. Nonlocal quantum gravity: A review. \emph{Int. J. Mod. Phys. D} \textbf{2017}, {\em 26}, 1730020.

\bibitem{effaction} Codello, A.; Percacci, R.; Rachwal, L.; Tonero, A. Computing the Effective Action with the Functional Renormalization Group. \emph{Eur. Phys. J. C} \textbf{2016}, {\em 76}, 226. 
\bibitem{effactionproc} Rachwal, L.; Codello, A.; Percacci, R. One-Loop Effective Action in Quantum Gravitation. \emph{Springer Proc.~Phys.} \textbf{2016}, {\em 170},~395--400.

\bibitem{modesto} Modesto, L. Super-renormalizable Quantum Gravity. \emph{Phys. Rev. D} \textbf{2012}, {\em 86}, 044005. 

\bibitem{as1} Weinberg, S. Ultraviolet divergences in quantum theories of gravitation. In \emph{General Relativity: An Einstein Centenary Survey}; Hawking, S.W., Israel, W., Eds.; Cambridge University Press: Cambridge, UK, 1979; pp. 790--831. 

\bibitem{as2} Reuter, M.; Saueressig, F. Renormalization group flow of quantum gravity in the Einstein--Hilbert truncation. \emph{Phys. Rev. D} \textbf{2002}, {\em 65}, 065016. 


\bibitem{as3} Lauscher, O.; Reuter, M. Is quantum Einstein gravity nonperturbatively renormalizable? \emph{Class. Quantum Gravity} \textbf{2002}, {\em 19}, 483.


\bibitem{as4} Litim, D.F. Fixed points of quantum gravity. \emph{Phys. Rev. Lett.} \textbf{2004}, {\em 92}, 201301.


\bibitem{kuzmin} Kuz'min, Y.V. The Convergent Nonlocal Gravitation. \emph{Sov. J. Nucl. Phys.} \textbf{1989}, {\em 50}, 1011--1014.  





%
\bibitem{fractional1} Calcagni, G. Multifractional theories: An unconventional review. \emph{J. High Energy Phys.} \textbf{2017}, {\em 2017}, 138. 

\bibitem{fractional2} Calcagni, G. Towards multifractional calculus. \emph{Front. Phys.} \textbf{2018}, {\em 6}, 58. 
\bibitem{fractional3} Calcagni, G. Multi-fractional spacetimes, asymptotic safety and Horava-Lifshitz gravity. \emph{Int. J. Mod. Phys. A} \textbf{2013}, {\em 28}, 1350092. 

\bibitem{highder}
Biswas, T.; Gerwick, E.; Koivisto, T.; Mazumdar, A. Towards singularity and ghost free theories of gravity. \emph{Phys. Rev. Lett.} \textbf{2012}, {\em 108}, 031101.


\bibitem{shapiroconf1} Shapiro, I.L.; Zheksenaev, A.G. Gauge dependence in higher derivative quantum gravity and the conformal anomaly problem. \emph{Phys. Lett. B} \textbf{1994}, {\em 324}, 286--292.


\bibitem{shapiroconf2} de Berredo-Peixoto, G.; Shapiro, I.L. Conformal quantum gravity with the Gauss--Bonnet term. \emph{Phys. Rev. D} \textbf{2004}, {\em 70}, 044024.

\bibitem{Eli1} 
Elizalde, E.; Odintsov, S.D.; Romeo, A. Manifestations of quantum gravity in scalar QED phenomena. 
\emph{\mbox{Phys. Rev. D}} \textbf{1995}, {\em 51}, 4250.


\bibitem{Eli2} 
Elizalde, E.; Odintsov, S.D.; Romeo, A. Improved effective potential in curved space-time and quantum matter, higher derivative gravity theory.
\emph{Phys. Rev. D} \textbf{1995}, {\em 51}, 1680. 

\bibitem{Eli3} 
Elizalde, E.; Zheksenaev, A.G.; Odintsov, S.D.; Shapiro, I.L. A Four-dimensional theory for quantum gravity with conformal and nonconformal explicit solutions. 
\emph{Class. Quantum Gravity} \textbf{1995}, {\em 12}, 1385. 

\bibitem{Eli4} 
Elizalde, E.; Odintsov, S.D.; Romeo, A. Renormalization group properties of higher derivative quantum gravity with matter in (4-epsilon)-dimensions. 
\emph{Nucl. Phys. B} \textbf{1996}, {\em 462}, 315--329. 

\bibitem{dilaton}
Codello, A.; D'Odorico, G.; Pagani, C.; Percacci, R. The Renormalization Group and Weyl-invariance. 
\emph{Class.~Quantum Gravity}\textbf{ 2013}, {\em 30}, 115015.

\bibitem{hollowood} Hollowood, T.J. \emph{Renormalization Group and Fixed Points: In Quantum Field Theory}; SpringerBriefs in Physics;  Springer: Berlin/Heidelberg, Germany, 2013.


\bibitem{Johansson:2018ues}
Johansson, H.; Mogull, G.; Teng, F. Unraveling conformal gravity amplitudes. \emph{arXiv} \textbf{2018}, arXiv:1806.05124.

\bibitem{Johansson:2017srf}
Johansson, H.; Nohle, J. Conformal Gravity from Gauge Theory.  \emph{arXiv} \textbf{2017}, 
arXiv:1707.02965.


\bibitem{FradkinT1} Fradkin, E.S.; Tseytlin, A.A. Conformal Supergravity. \emph{Phys. Rep.} \textbf{1985}, {\em 119}, 233--362.

\bibitem{FradkinT2} Fradkin, E.S.; Tseytlin, A.A. Instanton Zero Modes in addition, Beta Functions in Supergravities, Conformal Supergravity. \emph{Phys. Lett. B} \textbf{1984}, {\em 134B}, 307.



\bibitem{FradkinT4} Fradkin, E.S.; Tseytlin, A.A. Asymptotic Freedom in Extended Conformal Supergravities. \emph{Phys. Lett. B} \textbf{1982}, {\em 110B}, 117--122.
\bibitem{FradkinT5} Fradkin, E.S.; Tseytlin, A.A. One Loop Beta Function in Conformal Supergravities. \emph{Nucl. Phys. B} \textbf{1982}, {\em 203},~157--178.

\bibitem{nlsugra} Giaccari, S.; Modesto, L. Nonlocal supergravity. \emph{Phys. Rev. D} \textbf{2017}, {\em 96}, 066021. 


\bibitem{BVreport} Barvinsky, A.O.; Vilkovisky, G.A. The Generalized Schwinger-Dewitt Technique in Gauge Theories and
Quantum Gravity. \emph{Phys. Rep.} \textbf{1985}, \emph{119}, 1-74.

\bibitem{Cordova:2016emh} Cordova, C.; Dumitrescu, T.T.; Intriligator, K. Multiplets of Superconformal Symmetry in Diverse Dimensions. \emph{arXiv} \textbf{2016},  arXiv:1612.00809.
\bibitem{litimcft2} Bond, A.D.; Litim, D.F.; Vazquez, G.M.; Steudtner, T. UV conformal window for asymptotic safety. \emph{\mbox{Phys. Rev. D}} \textbf{2018}, {\em 97}, 036019. 


\bibitem{hooft1} t Hooft, G. Local conformal symmetry in black holes, standard model, and quantum gravity. \emph{Int. J. Mod. Phys. D} \textbf{2016}, {\em 26}, 1730006.

\bibitem{hooft2} t Hooft, G. Local conformal symmetry: The missing symmetry component for space and time. \emph{Int. J. Mod. Phys. D} \textbf{2015}, {\em 24}, 1543001.

\bibitem{hooft3} t Hooft, G. Spontaneous breakdown of local conformal invariance in quantum gravity. \emph{Les Houches Lect.~Notes} \textbf{2015}, {\em 97}, 209--253.

\bibitem{hooft4} t Hooft, G. Singularities, horizons, firewalls, and local conformal symmetry. \emph{arXiv} \textbf{2015}, arXiv:1511.04427.

\bibitem{mannheim1} Mannheim, P.D. Making the Case for Conformal Gravity. \emph{Found. Phys.} \textbf{2012}, {\em 42}, 388--420. 




\bibitem{mannheim2} Mannheim, P.D. Mass Generation, the Cosmological Constant Problem, Conformal Symmetry, and the Higgs Boson. \emph{Prog. Part. Nucl. Phys.} \textbf{2017}, {\em 94}, 125--183. 

\bibitem{mannheim3} Mannheim, P.D. Conformal Invariance and the Metrication of the Fundamental Forces. \emph{Int. J. Mod. Phys. D} \textbf{2016}, {\em 25}, 1644003. 

\bibitem{expjizba} Jizba, P.; Kleinert, H.; Scardigli, F. Inflationary cosmology from quantum Conformal Gravity. \emph{Eur. Phys. J. C} \textbf{2015}, {\em 75}, 245. 

\bibitem{exp1} Zhang, Q.; Modesto, L.; Bambi, C. A general study of regular and singular black hole solutions in Einstein's conformal gravity. \emph{Eur. Phys. J. C} \textbf{2018}, {\em 78}, 506. 

\bibitem{exp2} Chakrabarty, H.; Benavides-Gallego, C.A.; Bambi, C.; Modesto, L. Unattainable extended spacetime regions in conformal gravity. \emph{J. High Energy Phys.} \textbf{2018}, {\em 2018}, 13. 






\end{thebibliography}
\end{document}